\documentclass{aa}
\usepackage[varg]{txfonts}
\usepackage{natbib,graphicx}
\usepackage{mathtools}
\usepackage{subfig}
\usepackage{comment}
\usepackage{xcolor}
\usepackage{diagbox} 
\usepackage[squaren, Gray, cdot]{SIunits}
\usepackage[citecolor=blue, linkcolor=blue, urlcolor = black, colorlinks = true]{hyperref}
\usepackage{bigints}
\usepackage{lscape}
\usepackage{fancyvrb}

\RecustomVerbatimCommand{\VerbatimInput}{VerbatimInput}%
{fontsize=\footnotesize,
 frame=lines,  
 framesep=2em, 
 rulecolor=\color{gray},
 commandchars=\|\(\), 
 commentchar=*        
}

\begin{document}

\title{Detecting deep axisymmetric toroidal magnetic fields in stars}
\subtitle{The traditional approximation of rotation for differentially\\ rotating deep spherical shells with a general azimuthal magnetic field}
\author{H. Dhouib \inst{1}
\and S. Mathis \inst{2}
\and L. Bugnet \inst{3}
\and T. Van Reeth \inst{4}
\and C. Aerts \inst{4,5,6}
}

\institute{Université
de Paris, Université Paris-Saclay, CEA, CNRS, Astrophysique, Instrumentation et Modélisation Paris-Saclay,  F-91191, Gif-sur-Yvette, France\\\email{hachem.dhouib@cea.fr}
\and Université Paris-Saclay, Université
de Paris, CEA, CNRS, Astrophysique, Instrumentation et Modélisation Paris-Saclay,  F-91191, Gif-sur-Yvette,
\and Flatiron Institute, Simons Foundation, 162 Fifth Ave, New York, NY 10010, USA
\and Institute of Astronomy, KU Leuven, Celestijnenlaan 200D, 3001 Leuven, Belgium
\and Department of Astrophysics, IMAPP, Radboud University Nijmegen, PO Box 9010, 6500 GL Nijmegen, the Netherlands
\and Max Planck Institute for Astronomy, Koenigstuhl 17, 69117 Heidelberg, Germany}

\titlerunning{Detecting deep axisymmetric toroidal magnetic fields in stars}
\authorrunning{H. Dhouib et al.}

\abstract
{Asteroseismology has revealed small core-to-surface rotation contrasts in stars in the whole Hertzsprung-Russell diagram. This is the signature of strong transport of angular momentum (AM) in stellar interiors. One of the plausible candidates to efficiently carry AM is magnetic fields with various topologies that could be present in stellar radiative zones. Among them, strong axisymmetric azimuthal (toroidal) magnetic fields have received a lot of interest. Indeed, if they are subject to the so-called Tayler instability, the accompanying triggered Maxwell stresses can transport AM efficiently. In addition, the 
electromotive force induced by the fluctuations of magnetic and velocity fields could potentially sustain a dynamo action that leads to the regeneration of the initial strong axisymmetric azimuthal magnetic field.}
{The key question we aim to answer is whether we can detect signatures of these deep strong azimuthal magnetic fields. The only way to answer this question is asteroseismology, and the best laboratories of study are intermediate-mass and massive stars with external radiative envelopes. Most of these are rapid rotators during their main sequence. Therefore, we have to study stellar pulsations propagating in stably stratified, rotating, and potentially strongly magnetised radiative zones, namely  magneto-gravito-inertial (MGI) waves.}
{We generalise the traditional approximation of rotation (TAR) by simultaneously taking general axisymmetric differential rotation and azimuthal magnetic fields into account. Both the Coriolis acceleration and the Lorentz force are therefore treated in a non-perturbative way. Using this new formalism, we derive the asymptotic properties of MGI waves and their period spacings.}
{We find that toroidal magnetic fields induce a shift in the period spacings of gravity ($\rm g$) and Rossby ($\rm r$) modes. An equatorial azimuthal magnetic field with an amplitude of the order of $10^5\,\rm G$ leads to signatures that are detectable in period spacings for high-radial-order $\rm g$ and $\rm r$ modes in $\gamma$\,Doradus ($\gamma\,$Dor) and slowly pulsating B (SPB) stars. More complex hemispheric configurations are more difficult to observe, particularly when they are localised out of the propagation region of MGI modes, which can be localised in an equatorial belt.}
{The magnetic TAR, which takes into account toroidal magnetic fields in a non-perturbative way, is derived. This new formalism allows us to assess the effects of the magnetic field in $\gamma$\,Dor and SPB stars on $\rm g$ and $\rm r$ modes. We find that these effects should be detectable for equatorial fields thanks to modern space photometry using observations from \textit{Kepler}, TESS CVZ, and PLATO.}

\maketitle

\section{Introduction}

Space-based asteroseismology has made Eddington's dream `to see' inside stars a reality \citep[e.g.][]{GarciaBallot2019,aerts2021}. 
One of the major discoveries of the
\textit{Kepler} space mission \citep[][]{borucki2010, Howell2014} is the signature of a strong extraction of angular momentum (AM) operating in the radiative zones of stars of all types throughout their evolution, leading to weak core-to-surface rotation contrasts \citep[e.g.][]{Beck2012RG,Deheuvels2014RG,Kurtz2014,Triana2015B,Aerts2017,vanreeth2018,Li2020,Saio2021}. Understanding such efficient transport of momentum in stellar interiors, and the related chemical mixing, is one of the major unsolved questions of modern stellar astrophysics \citep{Dumont2021,Pedersen2021}. These mechanisms must be taken into account to get a complete and coherent understanding of stars, their evolution, age, and remnants, and the impact on their galactic environment \citep[e.g.][]{MaederMeynet2000,HegerSpruit2005,Suijsetal2008,Maeder2009,Bouchaudetal2020}. State-of-the-art stellar structure and evolution codes that take rotation and the meridional flows into account, as well as the shear instabilities they trigger, predict internal AM that is two orders of magnitude too high compared to asteroseismic observations \citep[][]{Eggenberger2012,Marques2013,Cantiello2014,ouazzani2019}.

Magnetic fields in stellar radiative zones are one of the plausible candidates to efficiently carry AM. The best laboratory to study these fields are early-type stars, which have an outer radiative envelope. This allows us to characterise the properties of the fields at the stellar surface \citep{Morel2014,Neiner2015,Wade2016} using ground-based high-resolution spectropolarimetric surveys. Two types of fields have been detected. On the one hand, 10\% of early-type stars host large-scale, generally high-amplitude, stable fields. Since they seem to present no correlations between their properties and stellar mass, age, and rotation \citep{Shultzetal2019}, these magnetic fields are supposed to have a fossil origin. This means that they result from the relaxation of fields generated by a dynamo during pre-main-sequence (PMS) convective phases when convective layers convert into stably stratified ones \citep[][]{Braithwaite2004,Duez2010,Arlt2014,Emeriau-ViardBrun2017}. 
As shown by \citet{Tayler1980}, \citet{Braithwaite2009}, and \citet{Duez2010b}, 
axisymmetric fields are only stable over long periods when they consist of
a poloidal and toroidal component
in the meridional plane and azimuthal direction, respectively,
since pure toroidal or poloidal fields are unstable in stellar radiative zones \citep[][]{Tayler1973,Markey1973}. On the other hand, the remaining 90\% of intermediate-mass and massive stars can host other types of fields detected as small-scale weak fields at the stellar surface \citep[][]{Lignieresetal2009,Petitetal2011,Blazereetal2016a,Blazereetal2016b}. Three possibilities have been suggested to explain their origin. First, they could be generated by a dynamo action in the thin sub-surface convective layer \citep{CantielloBraithwaite2019,JermynCantiello2020}. Second, they might result from a failed relaxation due to the action of the rapid rotation \citep{BraithwaiteCantiello2013}. Finally, they can be the signature of fields resulting from instabilities due to deep-seated axisymmetric toroidal magnetic fields at the stellar surface \citep[][]{Auriereetal2007,Gauratetal2015}.

Both types of fields are able to transport AM very efficiently \citep[e.g.][]{MestelWeiss1987,CharbonneauMacGregor1993,GoughMcIntyre1998,Spruit1999,Garaud2002,Spruit2002,MathisZahn2005,Strugareketal2011,Fulleretal2019, Petitdemange2021}. However, since the seminal work by \cite{Spruit2002}, a strong interest has been devoted to magnetic fields resulting from the Tayler instability of an axisymmetric toroidal magnetic field. \cite{Spruit2002} indeed suggested that such a field could trigger a dynamo action in radiative convectively stable layers and that the resulting magnetic torque allows a very efficient transport of AM \citep[e.g.][]{MaederMeynet2003,HegerSpruit2005,Eggenbergeretal2019}. The scenario of an efficient dynamo loop in stellar radiative zones has been strongly debated in the literature \cite[][]{Braithwaite2006,Zahnetal2007,Gellertetal2008,Gellertetal2011,Fulleretal2019}. Recent global 3D numerical simulations by \cite{Petitdemange2021} provide the first convincing proof of how such a dynamo can be driven by a combination of the action of differential rotation and its related shear instability and the action of the Tayler instability. In this work, we consider that a strong axisymmetric toroidal field is present in the radiative envelope but does 
not emerge at the stellar surface, implying that spectropolarimetry will not be able to detect and characterise it. The key question to answer is thus, given the importance of potentially unstable strong axisymmetric toroidal fields for stellar magnetism and rotation, how they can be detected and characterised.

Asteroseismology is the key to addressing this problem. More specifically, magneto-asteroseismology \citep{Neiner2015,Mathisetal2021} is the best diagnostic tool. As it stands, magneto-asteroseismology consists in searching for the characteristic signatures of magnetic field strengths and configurations in the observed frequency spectra of stellar oscillations. Although this method is still in its infancy, with only a few studies for the Sun \citep{GoodeThompson1992,KieferRoth2018} and early-type stars \citep{TakataShibahashi1993,ShibahashiAerts2000}, it has recently seen a strong development motivated by the simultaneous use of space-based asteroseismology and ground-based spectropolarimetry \citep{Briquetetal2012,Briquetetal2013,Neineretal2015,Neineretal2017,Buysschaert2018}. First, \cite{Prat2019} and \cite{Prat2020} studied the modification of the frequency spectrum of gravito-inertial modes by axisymmetric and inclined mixed (i.e.\ poloidal plus toroidal) dipolar fossil fields. Gravito-inertial modes are the oscillation modes that propagate in rotating stellar radiative zones under the combined action of the buoyancy and the restoring Coriolis forces. The authors treated the field as a perturbation as a first step. They showed how the behaviour of the so-called period spacing (i.e.\ the difference in period between two modes of the same degree and azimuthal order but of consecutive radial order as a function of the period of these modes) is modified by the presence of the field. They identified saw-tooth patterns that cannot be produced by chemical stratification or by differential rotation \citep{Miglio2008,vanreeth2018}. \cite{VanBeeck2020} made a systematic exploration of this magnetic signature along the evolution of intermediate-mass pulsating $\gamma$\,Doradus 
($\gamma\,$Dor hereafter) and slowly pulsating B (SPB) stars and showed that the magnetic signatures differ appreciably from those due to rotation alone. They found that these signatures are measurable for dipolar gravity mode oscillations in terminal-age main-sequence (TAMS) stars for a  magnetic field with a near-core strength larger than $10^5\,\rm G$. Next, \cite{Bugnetetal2021} and \cite{Loi2021na} studied the perturbation of mixed gravito-acoustic modes propagating in evolved low- and intermediate-mass stars by axisymmetric and non-axisymmetric inclined mixed dipolar fossil fields, respectively. They identified asymmetries in the frequency splittings induced by the field. Finally, \cite{Mathisetal2021} derived the asymptotic theory for the perturbation of low-frequency gravito-acoustic and gravito-inertial modes by such an axisymmetric field. This theory is in excellent agreement with direct numerical computations coupled to state-of-the-art stellar oscillation codes. The asymptotic expressions obtained for frequency splittings open the path to the potential inversion of the internal distribution of magnetic fields in stellar radiative zones.

All these previous studies have assumed that the magnetic field is a perturbation. This assumption becomes questionable for the high amplitudes connected with strong axisymmetric toroidal, initially fossil, fields that trigger Tayler instability and potential Tayler-Spruit-like dynamos. It thus becomes necessary to study the impact of moderate to strong amplitude fields on stellar pulsation modes. Predictions of this impact can then be used to detect and characterise the field. A first step has been achieved by \cite{Loi2020b,Loi2020a} and \cite{Loi2020m} in the case of gravity modes \citep[see also][]{Schatzman1993,Rudraiah1972a,Barnes1998,MacGregorRogers2011,Asai2016,Lee2018}. 

In this work, we study the case of gravito-inertial waves. In the presence of a strong magnetic field, these waves are driven by the buoyancy force, the Coriolis acceleration, and the Lorentz force. Therefore, they become magneto-gravito-inertial (MGI) waves \citep[e.g.][]{Braginskiy1967,Friedlander1987,Friedlander1989}. For this purpose, we propose generalising the traditional approximation of rotation (TAR) that was first introduced in geophysics \citep{eckart1960} to study the dynamics of the shallow Earth atmosphere and oceans. This formalism is intensively used in asteroseismology for the study of low-frequency waves propagating in strongly stratified zones \citep[e.g.][]{Bouabidetal2013,vanreeth2016,aerts2021}. In the hydrodynamical case, the vertical component of the Coriolis acceleration along the stratification direction can be neglected because it is dominated by
the buoyancy force. This leads to velocities that are mostly horizontal. Therefore, the 2D non-separable wave propagation equation in the general case \citep[e.g.][]{Dintrans1999} becomes separable \citep[e.g.][]{Bildsten1996,lee+saio1997}, as in the non-rotating case.\ It can be written as a combination of one equation that describes the wave's horizontal structure, named the Laplace tidal equation \citep{laplace1799}, and one radial Schr\"odinger-like equation for the radial propagation when assuming the Cowling approximation \citep{cowling1941}, in which the Eulerian perturbation of the gravitational potential is neglected. 

The TAR, in its standard version, relies mainly on three assumptions. First, the rotation is assumed to be uniform. Second, the star is assumed to be spherical, so the centrifugal acceleration is neglected, namely $\Omega \ll \Omega_{\mathrm{K}}$, where $\Omega_{\mathrm{K}} \equiv \sqrt{G M / R^{3}}$ is the Keplerian critical (breakup) angular velocity, and $G$, $M$, and $R$ are  the universal constant of gravity, the mass of the star, and the stellar radius, respectively. Finally, the magnetic field is not taken into account. However, given recent findings and observations, new theoretical developments of the standard TAR have emerged. The uniform rotation assumption was abandoned by \citet{OgilvieLin2004} and \citet{mathis2009}, who took the effects of general differential rotation into account. Moreover, the centrifugal acceleration was taken into account first for slightly deformed stars \citep{mathis+prat2019,Henneco2021} using a perturbative approach and then in the presence of strong deformation with uniform \citep{dhouib2021a} and differential \citep{dhouib2021b} rotation in a non-perturbative way. These recent papers on the generalised TAR formulations also treated the detectability and the signature of the centrifugal acceleration and  differential rotation on gravito-inertial modes from modern space photometric data.

\begin{figure}
    \centering
     \resizebox{\hsize}{!}{\includegraphics{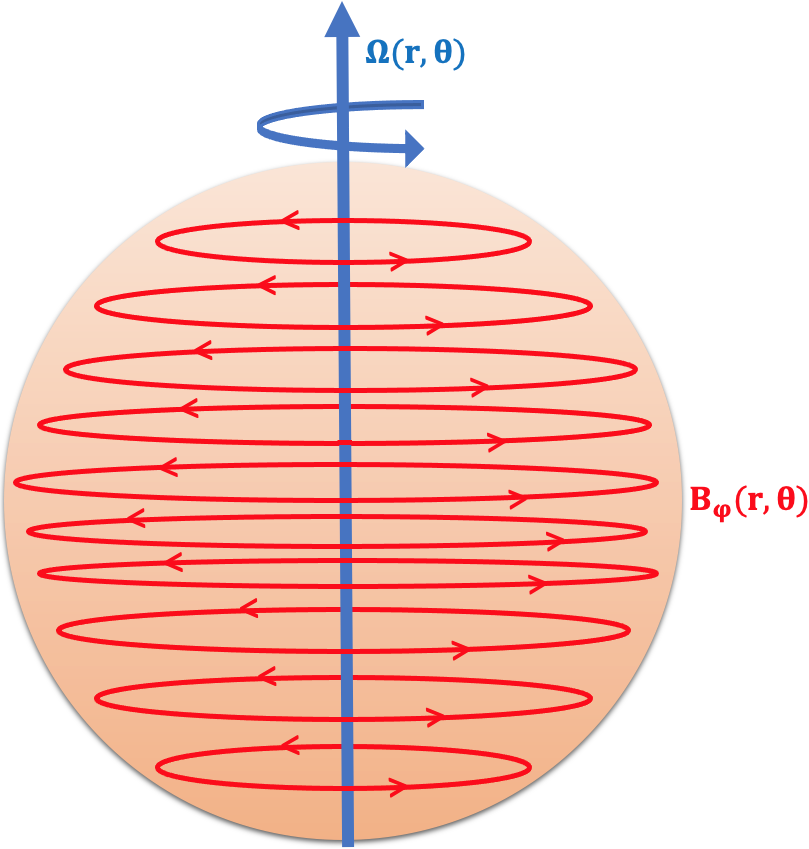}}
    \caption{Sketch of the set-up that we use to study the dynamics of MGI waves in rotating, magnetic, stably stratified stellar radiative zones.}
    \label{fig:schema}
\end{figure}

To introduce magnetic effects, \cite{Mathis&deBrye2011, Mathis&deBrye2012} took into account an axisymmetric toroidal magnetic field  for a uniform Alfvén frequency (the so-called \cite{Malkus1967} field) in a non-perturbative way and weak radial differential rotation in deep spherical shells to perform  a global study of MGI waves and the AM transport they trigger in stellar radiative regions \citep[a similar study was done by][]{Asai2015}. This approach allows for a comparison between the effects of rotation and magnetism. \cite{HengSpitkovsky2009} and \cite{Zaqarashvilietal2009} studied the cases of radial and azimuthal fields in shallow layers, respectively. Motivated by the observation of complex axisymmetric toroidal fields from numerical simulations \citep{Zahnetal2007,Jouveetal2020, Petitdemange2021}, our goal is to achieve a new generalisation of the TAR, going beyond the weak differential rotation approximation and the simplest magnetic azimuthal field configuration that corresponds to a uniform Alfvén frequency. To do so, we derive the magnetic TAR, which includes a general axisymmetric toroidal magnetic field (with a general Alfvén frequency), in a non-perturbative way and general differential rotation (cf. Fig.\;\ref{fig:schema}). The objective is to study the dynamics of MGI modes in the best asteroseismic targets available to detect them and characterise their frequency spectrum, namely $\gamma$\,Dor and SPB stars. In Sect.\;\ref{sect:MHD_equations} we present the linearised magnetohydrodynamics (MHD) system in differentially rotating fluids in the presence of a general toroidal axisymmetric magnetic field and the set of the adopted approximations. 

Section\;\ref{sect:Generalised_TAR} treats the derivation of
the magnetic TAR in this general configuration. Subsequently, we rewrite the MHD system in the form of a new magnetic Laplace tidal equation (MLTE) and deduce the asymptotic seismic diagnosis of MGI oscillation modes (Sect.\;\ref{sect:Dynamics_MGIWs}). As a proof of concept, we apply our formalism to typical models of $\gamma$\,Dor and SPB stars that host general axisymmetric toroidal magnetic fields as those observed in numerical simulations to study their seismic signature and their detectability (Sects.\;\ref{sect:results} and \ref{sect:hemispheric}).  Finally, we discuss our results, draw conclusions, and present perspectives and future applications of this work (Sect.\;\ref{sect:conclusion}).

\section{Propagation equation for MGI in differentially rotating  magnetic stars} \label{sect:MHD_equations}
\subsection{Ideal MHD equations system }
To study the dynamics of MGI waves in differentially rotating magnetic stars, we have to solve the ideal MHD non-dissipative dynamical equations system formed by the
induction equation, 
\begin{equation}\label{eq:induction_g}
    \partial_{t} \vec{B}=\vec{\nabla} \times(\vec{V} \times \vec{B})
,\end{equation}
with $\vec{\nabla}\cdot \vec{B}=0$, where we neglect the ohmic diffusion, the continuity equation,
\begin{equation} \label{eq:continuity_g}
    D_{t} \rho+\rho \vec{\nabla} \cdot \vec{V}=0,
\end{equation}
the momentum equation,
\begin{equation}\label{eq:momentum_g}
    D_{t} \vec{V}=-\frac{1}{\rho} \vec{\nabla} P-\vec{\nabla} \Phi+\frac{1}{\rho}\vec{F_{\mathcal{L}}},
\end{equation}
where the viscous friction is neglected, the energy transport equation in the adiabatic limit (i.e. without heat diffusion and viscous and ohmic heatings),
\begin{equation}\label{eq:energy_g}
\frac{1}{\Gamma_{1}} D_{t} \ln P-D_{t} \ln \rho=0,
\end{equation}
and Poisson's equation for the gravitational potential, $\Phi$, 
\begin{equation}\label{eq:poisson_g}
    \nabla^{2} \Phi=4 \pi G \rho,
\end{equation}
with $\rho$ the density, $P$ the pressure, and $G$ the universal gravitational constant.
We introduce the notations for the macroscopic velocity field, $\vec{V}$, the macroscopic magnetic field, $\vec{B}$, the Lagrangian derivative, 
$D_{t}=\partial_{t}+(\vec{V} \cdot \vec{\nabla})$ 
and consider the Lorentz force,
\begin{equation}\label{eq:Lorentz}
    \vec{F_{\mathcal{L}}} = \frac{1}{\mu}(\vec{\nabla} \times \vec{B}) \times \vec{B}.
\end{equation} 
Further, we denote
the magnetic permeability of the medium as $\mu$, while
$\Gamma_{1}=(\partial \ln P / \partial \ln \rho)_{S}$ ($S$ being the macroscopic entropy) is the adiabatic exponent. We can write the  macroscopic  velocity  field, $\vec{V}$, as  the sum of the large-scale azimuthal velocity, $\vec{V}_{0}$, associated with the differential rotation, $\Omega(r,\theta)$, and of the wave velocity, $\vec{u}^\prime$:
\begin{equation}
    \vec{V}(\vec{r}, t)=\vec{V}_{0}(r, \theta)+\vec{u}^\prime(\vec{r}, t),
\end{equation}
with $\vec{V}_{0}(r, \theta)=  r \sin \theta \, \Omega(r,\theta) \,  \vec{e}_{\varphi}$, where $t$ is the time, $\vec{r}$ is the position vector, and $(r,\theta,\varphi)$ are the usual spherical coordinates with their associated unit vector basis $(\vec{e}_r, \vec{e}_\theta, \vec{e}_\varphi)$.

\subsection{Magnetic field topology}
Generalising \cite{Mathis&deBrye2011,Mathis&deBrye2012}, we write the macroscopic magnetic field as
\begin{equation}\label{eq:magnetic_field}
    \vec{B}(\vec{r}, t)=\vec{B}_{0}^{\mathrm{T}}(r, \theta)+\vec{b}^\prime(\vec{r}, t),
\end{equation}
with 
\begin{equation}\label{eq:B_T}
    \vec{B}_{0}^{\mathrm{T}}(r, \theta)=\sqrt{\mu \rho_0} r \sin \theta \omega_{\mathrm{A}}(r, \theta) \,\vec{e}_{\varphi},
\end{equation}
a large-scale general axisymmetric toroidal (i.e. azimuthal) field  associated with the general Alfvén frequency $\omega_\mathrm{A}(r,\theta)$, $\rho_0$ the background hydrostatic density, and 
\begin{multline}\label{eq:b_prime}
    \vec{b}^\prime(\vec{r},t)=\sqrt{\mu \rho_0}\left[\omega_{\rm A} \partial_\varphi \vec{\xi}-\frac{r \sin{\theta}}{\sqrt{\rho_0}}\left(\vec{\xi}\cdot\vec{\nabla}\left(\sqrt{\rho_0}\omega_{\rm A}\right)\right)\vec{e}_\varphi\right]-\\  \left(\vec{\nabla}\cdot \vec{\xi}\right)\vec{B}_{0}^{\mathrm{T}},
\end{multline}
the wave-induced magnetic field. It is calculated using the induction equation (Eq.\;\ref{eq:induction_g}), where we introduced the  Lagrangian displacement of the wave, $\vec{\xi}$, defined as \citep{unno1989} 
\begin{equation}
    \vec{u}^\prime=\left(\partial_{t}+\Omega \partial_{\varphi}\right) \vec{\xi}-r \sin \theta\left(\vec{\xi} \cdot \vec{\nabla} \Omega\right) \vec{e}_{\varphi}.
\end{equation}
Using Eq.\;(\ref{eq:Lorentz}), we retrieve the expression of the Lorentz force,
\begin{equation}
    \vec{F}_{\mathcal{L}}(\vec{r}, t)=\vec{F}_{\mathcal{L} 0}(r, \theta)+{\vec{f}^{T}}_{\mathcal{L}}^\prime(\vec{r}, t)+{\vec{f}^{P}}_{\mathcal{L}}^\prime(\vec{r}, t) +{\vec{f}}_{\mathcal{L} 2}^\prime(\vec{r}, t),
\end{equation}
with
\begin{equation}\label{eq:lorentz_zero}
    \vec{F}_{\mathcal{L} 0}(r, \theta)=\frac{1}{\mu}\left(\vec{\nabla} \times \vec{B}_{0}^{T}\right) \times \vec{B}_{0}^{T},
\end{equation}
the zero-order part of the Lorentz force that perturbs the stellar hydrostatic balance,
\begin{equation}\label{eq:magnetic_tension}
    {\vec{f}^{T}}_{\mathcal{L}}^\prime(\vec{r}, t)=\frac{1}{\mu}\left[\left(\vec{B}_{0}^{T} \cdot \vec{\nabla}\right) \vec{b}^\prime+(\vec{b}^\prime \cdot \vec{\nabla}) \vec{B}_{0}^{T}\right],
\end{equation}
the wave magnetic tension force,
\begin{equation}
    {\vec{f}^{P}}_{\mathcal{L}}^\prime(\vec{r}, t)=-\frac{1}{\mu} \vec{\nabla}\left(\vec{B}_{0}^{T} \cdot \vec{b}^\prime\right),
\end{equation}
the wave magnetic pressure force, and 
\begin{equation}\label{eq:lorentz_second}
    {\vec{f}}_{\mathcal{L} 2}^\prime(\vec{r}, t)=\frac{1}{\mu}\left(\vec{b}^\prime \cdot \vec{\nabla}\right) \vec{b}^\prime,
\end{equation}
the second-order part of the Lorentz force.

\subsection{Linearised MHD equations}
We linearise the MHD system around the rotating magnetic steady-state. Each scalar field $X\equiv\{P, \rho, \Phi\}$ is expanded as the sum of its hydrostatic value $X_0$ and of the wave's associated fluctuation $X^\prime$:
\begin{equation}
X(\vec{r}, t)=X_0(r,\theta)+{X}^\prime(\vec{r}, t).
\end{equation}
We focus on first-order terms, implying that we neglect Eq.\;(\ref{eq:lorentz_second}).
In addition, we neglect the non-spherical character of the hydrostatic background due to the deformation associated with the centrifugal acceleration and the zero-order part of the Lorentz force (Eq.\;\ref{eq:lorentz_zero}) following \citet{Duez2010c}. This implies that the background is independent of $\theta$, so $X_0=X_0(r)$. This approximations is well justified for MGI waves propagating close to the convective core boundary in intermediate-mass stars \citep{Henneco2021,dhouib2021a,Duez2010c}.

We expand each vectorial field $(\vec{x}^\prime \equiv \{\vec{u}^\prime,\vec{b}^\prime\})$ and fluctuations of scalar quantities  $(X^\prime \equiv \{\rho^\prime, P^\prime, \Phi^\prime\})$ into discrete Fourier series both in time and in azimuth
\begin{gather}
    \vec{x}^\prime(\vec{r}, t)  \equiv \sum_{\omega^{\mathrm{in}}, m}\left\{\vec{x}(r, \theta) \exp (i\omega^{\mathrm{in}} t) \exp(-im \varphi)\right\},\\
    X^{\prime}(\vec{r}, t)  \equiv \sum_{\omega^{\mathrm{in}}, m}\left\{\widetilde{X}(r, \theta) \exp (i\omega^{\mathrm{in}} t) \exp(-im \varphi)\right\},
\end{gather}
where $m$ is the azimuthal order and $\omega^{\mathrm{in}}$ is the wave eigenfrequency in an inertial reference frame. In a differentially rotating region, the waves are Doppler shifted so we can define the wave frequency $\omega$ in the rotating reference frame  as
\begin{equation} \label{eq:doppler_shift}
    \omega(r, \theta) = \omega^{\mathrm{in}} - m\Omega(r, \theta) .
\end{equation}
This allows us to write the wave-induced magnetic field (Eq.\;\ref{eq:b_prime}) as
\begin{multline}
    \vec{b}(r,\theta)=\frac{\sqrt{\mu \rho_0}}{\omega}\left[-m\omega_{\rm A} \vec{u}+i r \sin{\theta} \left( \frac{1}{\sqrt{\rho_0}}\left(\vec{u}\cdot\vec{\nabla}\left(\sqrt{\rho_0}\omega_{\rm A}\right)\right)+\right.\right.\\\left. \left.    m \frac{\omega_{\rm A}}{\omega}\left(\vec{u}\cdot \vec{\nabla}\Omega\right) + \omega_{\rm A} \left(\vec{\nabla}\cdot \vec{u}-\frac{im}{r\sin{\theta}}u_\varphi\right)  \right)\vec{e}_{\varphi} \right],
\end{multline}
and the wave magnetic tension force (Eq.\;\ref{eq:magnetic_tension}) as
\begin{multline}
    {\vec{f}^{T}}_{\mathcal{L}}(r, \theta)= \rho_0 m \frac{\omega_{\rm A}^{2}}{\omega}\left[im \vec{u}-2 \vec{e}_{z} \times \vec{u}\right]+\\
    m r \sin{\theta} \rho_0\frac{\omega_{\rm A}^2 }{\omega} \left[ \left(\vec{\nabla}\cdot\vec{u}-\frac{im}{r \sin{\theta}}u_\varphi\right)  + \frac{m}{\omega} \left(\vec{u}\cdot\vec{\nabla}\Omega\right) \right] \vec{e}_\varphi-\\
    \frac{ i r \sin \theta \rho_0}{\omega}\left[m\Lambda_{\rm E} \left(\vec{u}\cdot \vec{\nabla}\Omega^2\right)+\frac{1}{\rho_0}\left(\vec{u} \cdot \vec{\nabla}\left( \rho_0 \omega_{\rm A}^{2}\right)\right) +
    \right.\\\left.
    2\omega_{\rm A}^{2}\left(\vec{\nabla}\cdot\vec{u}-\frac{im}{r \sin{\theta}}u_\varphi\right) \right] \vec{e}_{s},
\end{multline}
where $\vec{e}_z = \cos \theta\,\vec{e}_r - \sin \theta\,\vec{e}_\theta$ is the unit vector along the rotation axis, $\vec{e}_s=\sin \theta \,\vec{e}_r + \cos \theta \,\vec{e}_\theta$ is the radial unit vector of the cylindrical coordinates and $\vec{u}=\left(u_r,u_\theta,u_\varphi\right)$.
With this, we derive the linearised MHD system, where the continuity equation (Eq.\;\ref{eq:continuity_g}) becomes
\begin{equation}\label{eq:continuity_l}
    i\omega\widetilde{\rho}+\frac{1}{r^{2}} \partial_{r}\left(r^{2} \rho_0  u_{r}\right)+\frac{\rho_0}{r \sin{\theta}}\left[ \partial_{\theta}\left(\sin \theta u_{\theta}\right) - im u_{\varphi}\right]=0.
\end{equation}
We write the momentum equation (Eq.\;\ref{eq:momentum_g}) as

\begin{multline}\label{eq:momentum_l}
    i\frac{\mathcal{A}}{\omega} \vec{u} +\frac{\mathcal{B}}{\omega}  \vec{e}_z\times\vec{u} +\frac{r \sin{\theta}}{\omega}\left[i m \Lambda_{\rm E}\left(\vec{u}\cdot\vec{\nabla}\Omega^2\right)+ \phantom{\frac{r \sin{\theta}}{\omega}} \right.\\\left.\frac{i}{\rho_0}\left(\vec{u}\cdot\vec{\nabla}\left(\rho_0\omega_{\rm A}^2\right)\right)+2i\omega_{\rm A}^2\left(\vec{\nabla}\cdot\vec{u}-\frac{im}{r \sin{\theta}}u_\varphi\right)\right] \vec{e}_s+\\
    \frac{r\sin{\theta}}{\omega^2}\left[\mathcal{A}\left(\vec{u}\cdot\vec{\nabla}\Omega\right)+ m\omega_{\rm A}^2 \omega \left(\vec{\nabla}\cdot\vec{u}-\frac{im}{r \sin{\theta}}u_\varphi\right)\right]\vec{e}_\varphi=\\-\vec{\nabla} \widetilde{W}- \frac{\vec{\nabla}\rho_0}{\rho_0}\widetilde{W} +\frac{\widetilde{\rho}}{\rho_0} \vec{g_0} -\vec{\nabla}\widetilde{\Phi},
\end{multline}
with 
\begin{equation}
    \vec{g}_0=\vec{\nabla}P_0/\rho_0=-\vec{\nabla}\Phi_0+ \vec{\gamma_{\rm c}}
\end{equation}
the background effective gravity. We recover $\vec{\gamma_{\rm c}}=\frac{1}{2}\Omega^2\vec{\nabla}( r^2\sin^2{\theta})$ the centrifugal acceleration, introducing non-spherical perturbations of the hydrostatic background. They are neglected here as already argued before. We introduce
\begin{equation}
    \Lambda_{\mathrm{E}}(r,\theta)=\frac{\omega_{\mathrm{A}}^{2}}{\Omega \omega},
\end{equation}
the wave’s Elsasser  number, and 
\begin{equation}\label{eq:w_g}
    \widetilde{W}(r,\theta)= \frac{\widetilde{P}+\widetilde{P}_{m}}{\rho_0} + \widetilde{\Phi},
\end{equation}  
the sum of the normalised total pressure fluctuations (gas pressure fluctuation $\widetilde{P}$ and magnetic pressure fluctuation $ \widetilde{P}_{m} = \left(\vec{B}_{0}^{T} \cdot \vec{b}\right)/\mu $) and the gravitational potential fluctuation.
We introduce  the coefficients
\begin{equation}
    \mathcal{A}(r,\theta) =\omega^{2}-m^{2} \omega_{\rm A}^{2}
\end{equation}
and
\begin{equation}
     \mathcal{B}(r,\theta) =2\left(\Omega \omega+m \omega_{\rm A}^{2}\right).
\end{equation}
We rewrite the equation of energy (Eq.\;\ref{eq:energy_g}) in the adiabatic limit as 
\begin{equation}\label{eq:energy_l}
      i\omega \left(\frac{\widetilde{\rho}}{\rho_0}-\frac{\widetilde{P}}{\Gamma_1 P_0}\right)=\frac{N^{2}}{g_0} u_{r},
\end{equation}
where 
\begin{equation}
    N^2(r) = g_0\left(\frac{1}{\Gamma_{1}} \frac{\mathrm{d} \ln{P_0}}{\mathrm{d} r}- \frac{\mathrm{d} \ln{\rho_0}}{\mathrm{d} r}\right)
\end{equation}
is the squared Brunt–Väisälä frequency.
Finally, the Poisson equation (Eq.\;\ref{eq:poisson_g}) becomes
\begin{equation}\label{eq:poisson_l}
    \nabla^{2} \widetilde{\Phi}=4 \pi G \widetilde{\rho}.
\end{equation}
From now on, we adopt the \cite{cowling1941} approximation in which the wave's gravitational potential fluctuation $\widetilde{\Phi}$ is neglected. 
Thus, we do not solve Eq.\;(\ref{eq:poisson_l}) and neglect the term $\vec{\nabla}\widetilde{\Phi}$ in the momentum equation (Eq.\;\ref{eq:momentum_l}). Furthermore, because we study the dynamics of low-frequency MGI waves, we adopt the anelastic approximation in which magneto-acoustic waves are filtered out.
This means that the term $i\omega\widetilde{\rho}$ can be neglected in the continuity equation (Eq.\;\ref{eq:continuity_l})  and Eqs.\;(\ref{eq:momentum_l}) and\;(\ref{eq:energy_l}) can be simplified  by neglecting the terms $(\vec{\nabla}\rho_0/\rho_0)\widetilde{W}$ and $(1 / \Gamma_{1}) \widetilde{P} / P_0$, respectively.

\section{Magnetic TAR in the differentially rotating case} \label{sect:Generalised_TAR}
By adopting the approximations introduced in the previous section, we  write  the  three components of the momentum equation (Eq.\;\ref{eq:momentum_l}) as
\begin{multline}\label{eq:momentum_r}
        i \mathcal{A} u_{r}-\mathcal{B} \sin \theta u_{\varphi}+i r \sin ^{2} \theta\left[ m \Lambda_{E}\left( \vec{u} \cdot \vec{\nabla} \Omega^{2}\right)+\right.\\
        \left.\frac{1}{\rho_0}\left(\vec{u} \cdot \vec{\nabla}\left( \rho_0\omega_{\rm A}^{2}\right)\right)\right]=-\omega \partial_{r} \widetilde{W} +i N^2 u_r,
\end{multline}
\begin{multline}\label{eq:momentum_t}
    i \mathcal{A} u_{\theta}-\mathcal{B} \cos \theta u_{\varphi}+ir \sin \theta \cos \theta\left[m\Lambda_{E}\left( \vec{u} \cdot \vec{\nabla} \Omega^{2}\right)+\right.\\
    \left.\frac{1}{\rho_0}\left(\vec{u} \cdot \vec{\nabla}\left( \rho_0\omega_{\rm A}^{2}\right)\right)\right]=-\frac{\omega}{r} \partial_{\theta} \widetilde{W},
\end{multline}
\begin{multline}\label{eq:momentum_p}
    i \mathcal{A} u_{\varphi}+\mathcal{B} \cos \theta u_{\theta}+\mathcal{B}\sin \theta u_{r}+\frac{\mathcal{A}}{\omega} r \sin \theta\left(\vec{u} \cdot \vec{\nabla} \Omega\right)=\\
    \frac{i m \omega}{r \sin \theta}\widetilde{W}.
\end{multline}

The standard TAR imposes mainly two frequency hierarchies. The first one $2\Omega\ll N$ comes from the fact that we work in strongly stratified regions, which is the case for many stellar radiative zones. The second $\omega\ll N$ comes from focusing on the low-frequency regime.
We now build the generalised framework for the TAR in magnetic differentially rotating stars. In this case, in addition to the two abovementioned hierarchies, we also have $\mathcal{A}\ll N^2$ and $\mathcal{B}\ll N^2$ since $\omega_{\rm A}\ll N$. Therefore, the radial momentum equation (Eq.\;\ref{eq:momentum_r}) can be simplified to
\begin{equation}\label{eq:momentum_r_tar}
    N^2 u_r+ i\omega \partial_{r} \widetilde{W} =0.
\end{equation}
Moreover, the strong stratification implies that the radial components of the Coriolis force and the Lorentz force are dominated by the buoyancy force. From this, we obtain a mostly horizontal velocity field ($|u_r|\ll\{|u_\theta|,|u_\varphi|\}$). Therefore, the latitudinal momentum equation (Eq.\;\ref{eq:momentum_t}) can be reduced to
\begin{multline}\label{eq:momentum_t_tar}
    i \left(\mathcal{A}+\sin{\theta} \cos{\theta} \left(\Lambda_{\rm E}m\partial_\theta\Omega^2+\partial_\theta\omega_{\rm A}^2\right)\right)u_\theta-\mathcal{B}\cos{\theta}u_\varphi =\\-\frac{\omega}{r}\partial_\theta \widetilde{W}.
\end{multline}
Finally, the azimuthal momentum equation (Eq.\;\ref{eq:momentum_p}) can be rewritten, for the same reasons, as
\begin{equation}\label{eq:momentum_p_tar}
    \left(\mathcal{B}\cos{\theta}+\frac{\mathcal{A}}{\omega}\sin{\theta}\partial_\theta \Omega\right)u_\theta+ i\mathcal{A}u_\varphi=\frac{i m \omega}{r \sin{\theta}}\widetilde{W}.
\end{equation}

As in the non-magnetic case \citep{lee+saio1997, mathis2009, dhouib2021a, dhouib2021b}, we manage to decouple the radial and the horizontal dynamics. In the radial direction the buoyancy force balances the gradient of the total pressure fluctuations, while in the horizontal direction, the Coriolis acceleration and the Lorentz force tension counterbalance the horizontal gradient of the total pressure. Therefore, by solving the system (\ref{eq:momentum_r_tar})-(\ref{eq:momentum_p_tar}), we can express the velocity field as a function of the normalised pressure,
\begin{gather}
    u_r(r, x)=-i\frac{\omega}{N^2}\partial_r\widetilde{W}(r, x), \label{eq:u_r}\\
    u_\theta(r, x) = -i\frac{\omega}{\mathcal{A}}  \frac{1}{D_{\rm M}} \frac{1}{r} \frac{1}{\sqrt{1-x^2}} \left[\left(1-x^2\right)\partial_x+m\nu_{\rm M} x\right]\widetilde{W}(r, x),
    \label{eq:u_theta}
\end{gather}
\begin{multline}
    u_\varphi(r, x)= \frac{\omega}{\mathcal{A}}  \frac{1}{D_{\rm M}} \frac{1}{r} \frac{1}{\sqrt{1-x^2}} \left[\left(1-x^2\right)\left(\nu_{\rm M} x -\left(1-x^2\right) \frac{\partial_x\Omega}{\omega}\right)\partial_x + \right.\\\left.m - m \frac{1-x^2}{x} \frac{1}{\mathcal{A}}\left(m \Lambda_{\rm E} \partial_x\Omega^2 + \partial_x \omega^2_{\rm A} \right) \right]\widetilde{W}(r, x), \label{eq:u_phi}
\end{multline}
where $x=\cos{\theta}$ is the reduced latitudinal coordinate, 
\begin{equation}
    \nu_{\rm M}(r, x) = \frac{\mathcal{B}}{\mathcal{A}}=\nu \frac{\omega^2}{\mathcal{A}}+2m\frac{\omega_{\rm A}^2}{\mathcal{A}}
\end{equation}
is the magnetic spin parameter,
\begin{equation}
    \nu (r, x) = \frac{2\Omega}{\omega}
\end{equation}
is the standard rotation spin parameter, and
\begin{equation}
    D_{\rm M} (r, x) = 1- \nu_{\rm M}^2 x^2+\left(1-x^2\right)\frac{x}{\mathcal{A}}\left[\partial_x\Omega^2 -\partial_x\omega_{\rm A}^2\right]
\end{equation}
is a coefficient depending on the rotation rate, on the Alfvén frequency, and on their latitudinal gradient. We note that these equations are singular at $\omega=0$, $\mathcal{A}=0$, $x=\{\pm 1$, 0\}, and $D_{\rm M}=0$. The first two singularities are potentially the hydrodynamic and hydromagnetic critical layers, where 
the waves can be damped \citep{booker&bretherton1967,Rudraiah1972a}. The first singularity $\omega=0$ corresponds to a corotation resonance, which is present in the hydrodynamic case \citep{booker&bretherton1967, Watts2003, Alvan2013, Astoul2021}. The second singularity, which appears only in the magnetic case, adds two supplementary critical layers, $\omega= \pm m\omega_{\rm A}$, thus called hereafter magnetic critical layers. They can lead to MGI vertical trapping \citep{Rudraiah1972a,Schatzman1993,Barnes1998,Rogers2010,MacGregorRogers2011,Mathis&deBrye2011}. This is a situation similar to those studied by \cite{Fuller2015} and \cite{Lecoanet2017} in which MGI waves are converted into Alfv\'en waves above a given critical field strength, preventing them from forming a standing oscillation mode. The difference is that we do not take into account any vertical (poloidal) magnetic field in our work. The third singularity arises at the poles and at the equator. Finally, $D_{\rm M}=0$ is equivalent to a Lindblad resonance in differentially
rotating disks \citep[e.g.][]{Goldreich1979} but it is only an apparent singularity \citep{OgilvieLin2004}. More details on this turning point for the propagation of waves can be found in Sect.\;\ref{sect:hough_eigen}.

The above new theory generalises \cite{mathis2009} by taking into account axisymmetric toroidal magnetic fields and \cite{Mathis&deBrye2011} by accounting for general axisymmetric Alfv\'en and rotation frequency distributions $\omega_{\rm A}\left(r,\theta\right)$ and $\Omega\left(r,\theta\right)$ instead of a uniform $\omega_{\rm A}$ and a solid-body rotation $\Omega_{\rm s}$. 
Our derivations complete the pioneering general work by \cite{Friedlander1987,Friedlander1989} in geophysics, who studied MGI waves with $\left(\omega_{\rm A}\left(r,\theta\right),\Omega_{\rm s}\right)$ and $\left(\omega_{\rm A}\left(r,\theta\right),\Omega_{\rm}\left(r,\theta\right)\right)$, respectively. 
Compared to these two earlier studies, we focused specifically on the stratification-dominated regime relevant in stellar interiors for which the TAR can be built and applied. It is also noteworthy that we treated the case of deep spherical shells, while many previous studies focused on the case of thin layers \citep[e.g.][]{Zaqarashvilietal2009,HengSpitkovsky2009}.

\section{Dynamics of low-frequency MGI waves}\label{sect:Dynamics_MGIWs}
In order to study the dynamics of low-frequency MGI waves, we derive the MLTE for the normalised pressure. This equation allows us to build seismic diagnostics of MGI waves by computing the asymptotic frequencies as in \cite{vanreeth2018}, \cite{Henneco2021}, and \cite{dhouib2021a, dhouib2021b}.

\subsection{JWKB approximation}
Following \cite{mathis2009}, we use the 2D Jeffreys-Wentzel-Kramers-Brillouin (JWKB) approximation \citep{froman1965} since we focus on rapidly oscillating waves along the radial direction. In this case, we write the spatial structure of the waves as
\begin{align}\label{eq:jwkb_w}
\widetilde{W}(r, \theta)&=\sum_{k}\left\{w_{\omega^{\mathrm{in}} k m}(r, \theta) \frac{A_{\omega^{\mathrm{in}} k m}}{k_{V ; \omega^{\mathrm{in}} k m}^{1 / 2}} \exp \left[i \int^{r} k_{V ; \omega^{\mathrm{in}} k m} \mathrm{d} r\right]\right\},
\\u_{j}(r, \theta)&=\sum_{k}\left\{\hat{u}^j_{\omega^{\mathrm{in}} k m}(r, \theta) \frac{A_{\omega^{\mathrm{in}} k m}}{k_{V ; \omega^{\mathrm{in}} k m}^{1 / 2}} \exp \left[i \int^{r} k_{V;\omega^{\mathrm{in}} k m} \mathrm{d} r \right]\right\},\label{eq:jwkb_u}
\end{align}
with $j \equiv\{r, \theta, \varphi\}$, $k$ the index of a latitudinal eigenmode, and $A_{\omega^{\mathrm{in}} k m}$ the amplitude of the wave. This approximation allows us to rewrite the radial velocity (Eq.\;\ref{eq:u_r}) as 
\begin{equation}\label{eq:u_r_final}
        \hat{u}^r_{\omega^{\mathrm{in}} k m}(r, x) =  \frac{r^2 k_{V;\omega^{\mathrm{in}} k m}(r)\omega_{ k m}(r, x)}{N^2(r)} w_{\omega^{\mathrm{in}} k m}(r, x),
\end{equation}
so, the continuity equation (Eq.\;\ref{eq:continuity_l}) can be written as
\begin{multline}
    \label{eq:continuty_jwkb}
      \frac{1}{\omega_{ k m}}\frac{i}{r \sin{\theta}}\left[ \partial_{\theta}\left(\sin \theta \hat{u}^\theta_{\omega^{\mathrm{in}} k m}\right) - im \hat{u}^\varphi_{\omega^{\mathrm{in}} k m}\right]=\\ \frac{r^2 k_{V;\omega^{\mathrm{in}} k m}^2}{N^2} w_{\omega^{\mathrm{in}} k m}(r, x) ,
\end{multline}
where we neglect the terms $\displaystyle{ 2\frac{\rho_0}{r}  u_{r}}$ and $\displaystyle{\mathrm{d}_r\rho_0 u_r}$ in front of the dominant term $\rho_0\partial_r u_r$ since within the JWKB approximation the highest order derivative term in the radial direction is the dominant one.

\subsection{Magnetic Laplace tidal equation (MLTE)} \label{subsect:glte}
By inserting the horizontal components of the velocity (Eqs.\;\ref{eq:u_theta} and\;\ref{eq:u_phi}) in the 
left-hand side
of the continuity equation (Eq.\;\ref{eq:continuty_jwkb}), we get the magnetic Laplace tidal operator (MLTO),
\begin{multline}\label{eq:mlto_general}
    \widetilde{\mathcal{L}}^{\rm{magn.}}_{\omega^{\mathrm{in}} m} = \frac{1}{\omega} \partial_x \left[\frac{\omega}{\mathcal{A}}\frac{1-x^2}{D_{\rm M}}\partial_x\right] + \frac{m}{\omega\mathcal{A}}\partial_x\Omega \frac{1-x^2}{D_{\rm M}} \partial_x +
    \frac{m}{\omega} \partial_x\left(\frac{\omega \nu_{\rm M} x}{\mathcal{A}D_{\rm M}}\right)\\-m^2\frac{1}{\mathcal{A}D_{\rm M}\left(1-x^2\right)} + \frac{m^2}{\mathcal{A}^2}\frac{x}{D_{\rm M}}\left(m \Lambda_{\rm E} \partial_x \Omega^2 + \partial_x \omega_{\rm A}^2\right),
\end{multline}
such that
\begin{equation}\label{eq:mlte}
    \widetilde{\mathcal{L}}^{\rm{magn.}}_{\omega^{\mathrm{in}} m} \left[w^{\rm{magn.}}_{\omega^{\mathrm{in}} k m}\right] = -\lambda^{\rm{magn.}}_{\omega^{\mathrm{in}} k m} \left(r\right) w^{\rm{magn.}}_{\omega^{\mathrm{in}} k m}\left(r, \theta\right),
\end{equation}
where $w_{\omega^{\mathrm{in}} k m}$ are the magnetic generalised Hough functions and $\lambda_{\omega^{\mathrm{in}} k m}$ are the eigenvalues (with dimension
s$^2$) linked to the radial wave vector via the following dispersion relation deduced from the left-hand side of  Eq.\;(\ref{eq:continuty_jwkb}),
\begin{equation}\label{eq:dispersion}
    k_{V ; \omega^{\mathrm{in}} k m}^{2}\left(r\right)=\frac{N^2\left(r\right) \lambda_{\omega^{\mathrm{in}} k m}^{\rm{magn.}}\left(r\right)}{r^2}.
\end{equation}
Contrary to the standard case where we consider uniformly rotating non-magnetic stars, the eigenvalues and the Hough functions are no longer independent of the radius. We also remark that the MLTO is invariant with respect to the sign of the Alfvén frequency $\omega_{\rm A}$.
With the dispersion relation (Eq.\;\ref{eq:dispersion}), we differentiate two wave regimes. The first one is the propagative one ($k_{V ; \omega^{\mathrm{in}} k m}^{2}>0$) obtained in the stably stratified region ($N^2>0$) when  $\lambda_{\omega^{\mathrm{in}} k m}>0$ or in the convective region ($N^2<0$) when $\lambda_{\omega^{\mathrm{in}} k m}<0$. The second regime is the evanescent one ($k_{V ; \omega^{\mathrm{in}} k m}^{2}<0$) obtained in the stably stratified region when  $\lambda_{\omega^{\mathrm{in}} k m}<0$ or in the convective zone when $\lambda_{\omega^{\mathrm{in}} k m}>0$. The magnetic TAR developed here is only applicable in  stably stratified regions and it assumes adiabatic oscillations, that is, we suppose that $N^2=0$ in convective zones. Thus, we focus only on propagative waves in stably stratified regions.
In the non-magnetic case ($\omega_{\rm A}=0$), the MLTO (Eq.\;\ref{eq:mlto_general}) reduces to the hydrodynamic generalised Laplace tidal operator in the case of differential rotation derived by \cite{mathis2009},
\begin{multline}
    \widetilde{\mathcal{L}}_{\omega^{\mathrm{in}} m} = \frac{1}{\omega} \partial_x \left[\frac{1}{\omega}\frac{1-x^2}{\Tilde{D}}\partial_x\right] + \frac{m }{\omega^3}\partial_x\Omega \frac{1-x^2}{\Tilde{D}} \partial_x +\\
    \frac{m}{\omega}\partial_x\left(\frac{\nu x}{\omega \Tilde{D}}\right)-\frac{m^2}{\Tilde{D}\omega^2\left(1-x^2\right)},
\end{multline}
with
\begin{equation}
    \Tilde{D} = D_{\rm M}\vert_{\omega_{\rm A}=0} = 1- \nu^2 x^2+\nu x \left(1-x^2\right)  \frac{\partial_x\Omega}{\omega}.
\end{equation}

\subsection{Asymptotic seismic diagnosis}
The dispersion relation (Eq.\;\ref{eq:dispersion}) is not yet suitable for asteroseismic diagnosis because the radial wave number is not explicitly expressed as a function of the wave frequency. However, the eigenvalues  $\lambda_{\omega^{\mathrm{in}} k m}$ can be expressed as a function of the wave frequency and the dimensionless eigenvalues. 
In this case, it is necessary to assume that the local Doppler-shifted wave frequency $\omega$ only depends on the radius.

\subsubsection{Approximation on the differential rotation profile}
As in \cite{dhouib2021b}, we have to perform  a partial separation between the radial and latitudinal variables in the radial velocity  (Eq.\;\ref{eq:u_r_final}) to be able to derive the dimensionless MLTO. In fact, the wave frequency in the rotating frame  (Eq.\;\ref{eq:doppler_shift}) depends on $r$ and $\theta$. This dependence comes from the differential rotation $\Omega(r,\theta)$.
Therefore, we assume that the rotation rate $\Omega$ depends mainly on the radius $\Omega(r, \theta)\approx \Omega(r)$ to obtain a wave frequency that only depends on the radial coordinate $\omega(r, \theta)\approx \omega(r)$. In that case, the dimensionless MLTE can be expressed as
\begin{align}
    \mathcal{L}^{\rm{magn.}}_{\omega^{\mathrm{in}} m} \left[w^{\rm{magn.}}_{\omega^{\mathrm{in}} k m} (r,\theta)\right] &= \omega^2(r)\widetilde{\mathcal{L}}^{\rm{magn.}}_{\omega^{\mathrm{in}} m} \left[w^{\rm{magn.}}_{\omega^{\mathrm{in}} k m}(r,\theta)\right]\nonumber\\
    &= -\Lambda^{\rm{magn.}}_{\omega^{\mathrm{in}} k m}(r) w^{\rm{magn.}}_{\omega^{\mathrm{in}} k m}(r,\theta),
    \label{eq:mlto_general2}
\end{align}
with 
\begin{multline}
    \mathcal{L}^{\rm{magn.}}_{\omega^{\mathrm{in}} m} = \omega^2 \partial_x \left[\frac{1}{\mathcal{A}}\frac{1-x^2}{D_{\rm M}}\partial_x\right] +
    m\omega^2 \partial_x\left(\frac{ \nu_{\rm M} x}{\mathcal{A}D_{\rm M}}\right)\\-m^2\frac{\omega^2}{\mathcal{A}D_{\rm M}\left(1-x^2\right)} + m^2 \frac{\omega^2}{\mathcal{A}^2}\frac{x}{D_{\rm M}} \partial_x \omega_{\rm A}^2
\end{multline}
and
\begin{equation}
    \Lambda^{\rm{magn.}}_{\omega^{\mathrm{in}} k m} (r) = \omega^2(r) \lambda^{\rm{magn.}}_{\omega^{\mathrm{in}} k m}(r).
\end{equation}
In this way, the dispersion relation can be rewritten as
\begin{equation}\label{eq:dispersion2}
    k_{V ; \omega^{\mathrm{in}} k m}^{2}=\frac{N^2 \Lambda_{\omega^{\mathrm{in}} k m}^{\rm{magn.}}}{r^2\omega^2}.
\end{equation}
The normalisation used here to obtain the  dimensionless MLTE is different from the one used in \cite{Mathis&deBrye2011}. These authors normalise their tidal operator with $\mathcal{A}$ instead of $\omega^2$, since $\mathcal{A}$ is independent of $\theta$ in their study (uniform angular velocity and Alfvén frequency). Here, $\mathcal{A}$ depends on $r$ and $\theta$, so we design a new normalisation to be as general as possible in the derivation of the  dispersion relation (Eq.\;\ref{eq:dispersion2}).
We achieve this by performing an adequate partial variable separation. 
In this way, no constraints need to be placed on the 2D Alfvén frequency profile and on the corresponding field configuration to derive seismic diagnosis.

\subsubsection{Asymptotic frequency of low-frequency MGI waves}
Following \cite{vanreeth2018}, \cite{mathis+prat2019}, and \cite{dhouib2021a,dhouib2021b}, we derive the eigenfrequencies of low-frequency MGI waves by applying the radial quantisation \citep{unno1989, gough1993, Christensen1997},
\begin{equation}\label{eq:quantisation}
    \int_{r_{1}}^{r_{2}} k_{V ; \omega^{\mathrm{in}} n k m} \mathrm{d} r=(n+1 / 2) \pi,
\end{equation}
where $r_1$ and $r_2$ are the turning points of the 
Brunt–Väisälä frequency $N$ and $n$ is the radial order. 
At this point, no further analytical work can be done without further assumptions. Therefore, we have to solve Eq.\;(\ref{eq:quantisation}) numerically in order to retrieve the asymptotic frequency values.

\section{Axisymmetric azimuthal magnetic field in rotating intermediate-mass main-sequence stars}\label{sect:results}
As a proof of concept, we apply the magnetic TAR to representative stellar models of typical $\gamma$\,Dor and SPB stars during their main-sequence (MS) evolution because they are gravity or gravito-inertial wave pulsators allowing us to probe successfully mixing processes \citep{Degroote2010, Pedersen2021}, rotation \citep{vanreeth2016,vanreeth2018, Ouazzani2017,Papics2017, Aerts2017, aerts2019,Ouazzani2020,Saio2021} including centrifugal effects \citep{Henneco2021,dhouib2021a}, and potentially magnetic fields \citep{Prat2019,Prat2020, VanBeeck2020}.

\subsection{Studied $\gamma$\,Dor and SPB stars} \label{sect:mesa_setup}
\begin{figure}[htp]
\centering
\includegraphics[width=1\linewidth]{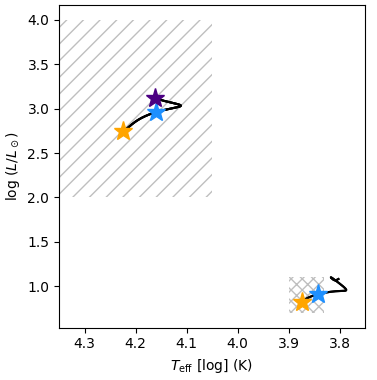}
\caption{Main-sequence tracks of a $M=1.6{\rm M}_\odot$ $\gamma$\,Dor star and a $M=5{\rm M}_\odot$ SPB star in the Hertzsprung-Russell diagram. The crossed (respectively\ hatched) area 
roughly indicates the region where $\gamma$\,Dor (respectively SPB) stars are expected to host gravity modes \citep[][Appendix\,A]{aerts2010}.
Yellow, blue, and purple stars indicate the chosen models at ZAMS, mid-MS, and TAMS, respectively. }
\label{fig:HR}
\end{figure}

We use the stellar evolution code Modules for Experiments in Stellar Astrophysics \citep[MESA;][]{Paxton2011, Paxton2013, Paxton2015, Paxton2018, Paxton2019} to compute non-rotating, non-magnetic stellar models for typical parameters used 
in asteroseismic modelling of $\gamma$\,Dor and SPB stars, following
\citet{vanreeth2016, Papics2017, Buysschaert2018}. We refer to the controls section of the MESA inlist in Appendix~\ref{sec:MESA_inlist}.
We select $1.6\mathrm{M}_{\odot}$ ($\gamma$\,Dor star) and $5\mathrm{M}_{\odot}$ (SPB star) models near the zero-age main sequence (ZAMS),
near the middle of the main sequence (mid-MS),
and for the SPB model also near the TAMS.
With this choice, we ensure that the models occur in areas in the Hertzsprung-Russell diagram where gravity modes are expected to occur. 
We schematically show this region in Fig.\;\ref{fig:HR}, following Appendix\,A in 
\citet{aerts2010}.

\subsection{Studied magnetic topology
} \label{sec:field_topo}
\begin{figure*}[htp]
\centering
\includegraphics[width=0.95\textwidth]{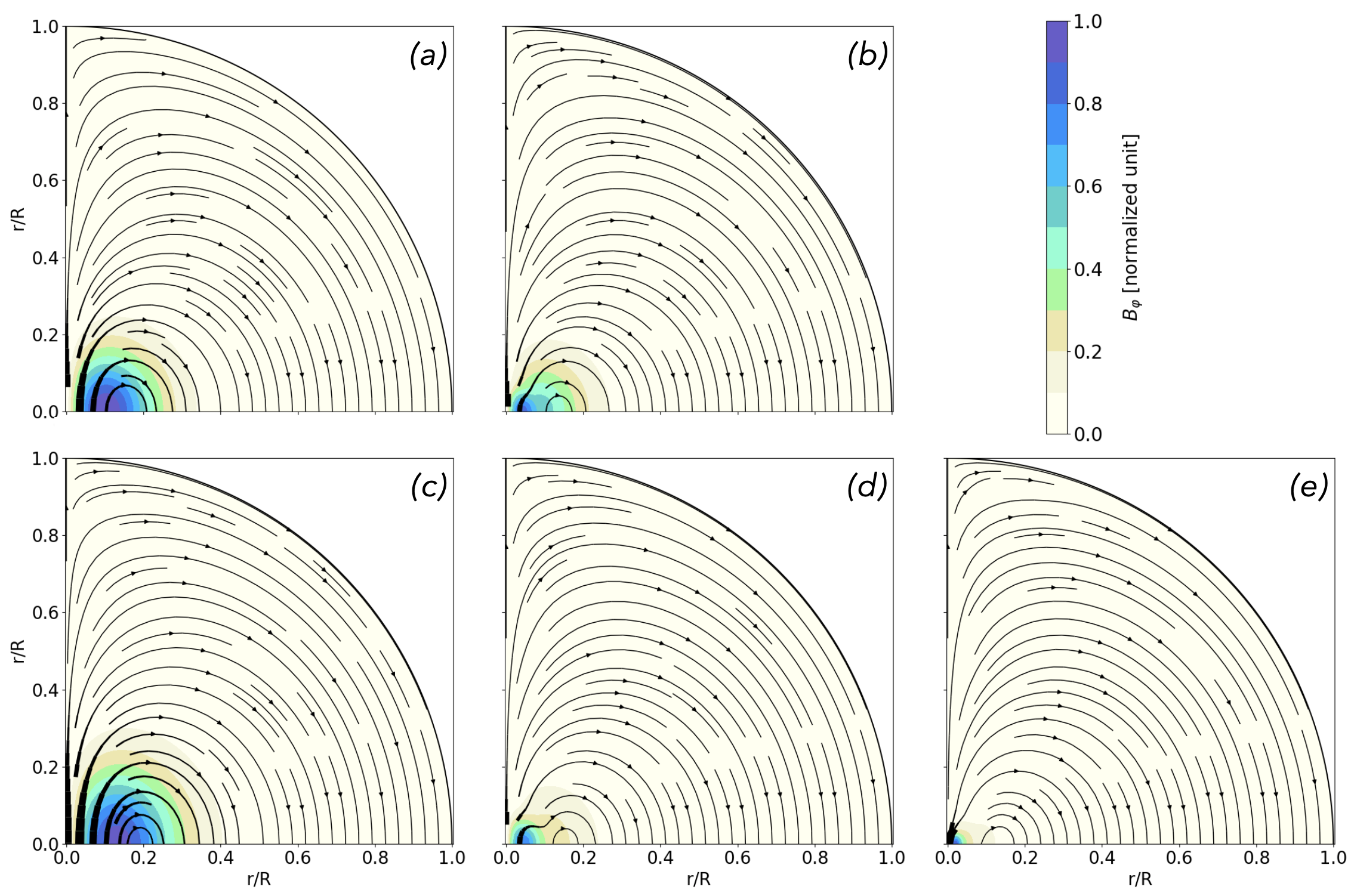}
\caption{Magnetic field configurations:  inside the $1.6{\rm M}_\odot$ $\gamma$\,Dor star near ZAMS \textsl{(a)}, at mid-MS \textsl{(b)}, inside the $5{\rm M}_\odot$ SPB near ZAMS \textsl{(c)}, at mid-MS \textsl{(d)}, and near TAMS \textsl{(e)}. Values are normalised by the maximum field amplitude, and mixed  poloidal (black lines) and toroidal (colour scale) magnetic fields are modelled using the formalism of \cite{Duez2010} (see our Eq.\;\eqref{eq:compu}). We focus here on the effects of the toroidal component as discussed in the text.}
\label{fig:mag_topo}
\end{figure*}

Intermediate-mass stars are descended from fully convective PMS stars. Such a convective body is known to generate and host dynamo-generated fields that might remain in the radiative envelope once the convection ends \citep[e.g.][]{Mathis2010b,Arlt2014,  Neiner2015, Emeriau-ViardBrun2017}. \cite{Braithwaite2004} simulated the relaxation of stochastic seed fields in a non-rotating radiative sphere. This can be used to study the relaxation of dynamo-generated fields once the plasma becomes stably stratified along the PMS \citep{Alecian2013}. From these simulations, we deduce that stochastic dynamo fields placed into a stably stratified region might relax into large-scale, mixed poloidal and toroidal magnetic field configurations. The stability of such mixed poloidal and toroidal fields has also been investigated. Purely toroidal and purely poloidal magnetic configurations are highly unstable \citep[e.g.][]{Tayler1973, Markey1973, Braithwaite2006, Braithwaite2007}. However, the combination of toroidal and poloidal components ensures the stability of the field \citep{Tayler1980}. The topology and energy of such a mixed (poloidal and toroidal) field has been investigated numerically \citep{Braithwaite2008a, Duez2010b}, and semi-analytically by \cite{Akgun2013}. \cite{Duez2010} provided a semi-analytic description of such stable mixed toroidal and poloidal fields in the non-rotating axisymmetric case:
\begin{equation}
    \vec{B}(r,\theta)=
    \displaystyle
    \frac{1}{r \sin{\theta}}\left( \vec{\nabla} \psi(r,\theta) \wedge \vec{e_\varphi} + \lambda \frac{\psi(r,\theta)}{R} \vec{e_\varphi}\right), 
    \label{eq:compu}
\end{equation}
\noindent where $\psi$ is the magnetic stream function satisfying 
\begin{equation}
    \psi(r,\theta)=\mu_0 \alpha \lambda \frac{A(r)}{R}\sin^2{\theta}\, ,
\end{equation} with $\mu_0$ the vacuum magnetic permeability, $\alpha$ a normalisation constant fixed by the chosen magnetic-field amplitude, $\lambda$ the eigenvalue of the problem to be determined, $R$ the radius of the star, and 
\begin{multline}
A(r)=-r j_1\left(\lambda \frac{r}{R} \right) \int_r^{R} y_1\left(\lambda \frac{x}{R} \right)\rho_0 x^3 \textrm{d}x\\
-r y_1\left(\lambda \frac{r}{R} \right) \int_0^r j_1\left(\lambda \frac{x}{R} \right)\rho_0 x^3 \textrm{d}x,
\label{eq:A}
\end{multline}
\noindent with $j_1$ ($y_1$) the first-order spherical Bessel functions of the first (second) kind \citep{Abramowitz1972}. 
\cite{Duez2010} derived lowest energy confined equilibrium states with $\lambda$ the smallest positive eigenvalue such that $\vec{B}=0$ at the stellar surface. Following \cite{Prat2019,Prat2020}, \cite{VanBeeck2020}, \cite{Bugnetetal2021}, and \cite{Mathisetal2021}, we use them to study the ability of asteroseismology to probe deep magnetic fields in stars. 
The resulting magnetic field configurations along the evolution of $\gamma$\,Dor and SPB stars are represented in Fig.~\ref{fig:mag_topo}.

In our study, we focus on the non-perturbative effect of axisymmetric toroidal fields on MGI waves computed within the TAR. We therefore chose here to extract the toroidal configuration from the mixed configuration defined from Eq.\;(\ref{eq:compu}) with the following correspondences to match the \cite{Duez2010} formalism:
\begin{equation}\label{eq:B_duez}
    \vec{B}=B_0\left[0,0, \tilde{b}_\varphi(r) \sin{\theta}\right]\, ,
\end{equation}
\noindent with 
\begin{equation}
    B_0=\frac{\mu_0 \alpha \lambda}{R}
    \end{equation}
    \noindent and
    \begin{equation}
    \tilde{b}_\varphi(r)=\frac{\lambda A}{r R}. \label{eq:bphi}
\end{equation}
If the differential rotation remains weak the field will remain stable and fossil-like. However, if a sufficient radial differential rotation builds up, its shear will wind up the poloidal component of the field into a supplementary toroidal field \citep{Auriereetal2007, Gauratetal2015, Jouveetal2020} that can become unstable because of the Tayler instability. If an efficient dynamo loop takes place, as first proposed by \cite{Spruit2002} and identified by \cite{Petitdemange2021}, the initial toroidal field can be regenerated. In \cite{Petitdemange2021}, this leads to a deep confined equatorial toroidal field that can be mimicked as a first rough approximation by Eq.\;(\ref{eq:B_duez}) even if $\tilde{b}_\varphi(r)$ will be different in this case because it is driven by the physics of the dynamo and not of the initial MHD relaxation forming the seed fossil field. More complex latitudinal dependences will be studied specifically in Sect.\;\ref{sect:hemispheric}.

\begin{figure}
    \centering
     \resizebox{\hsize}{!}{\includegraphics{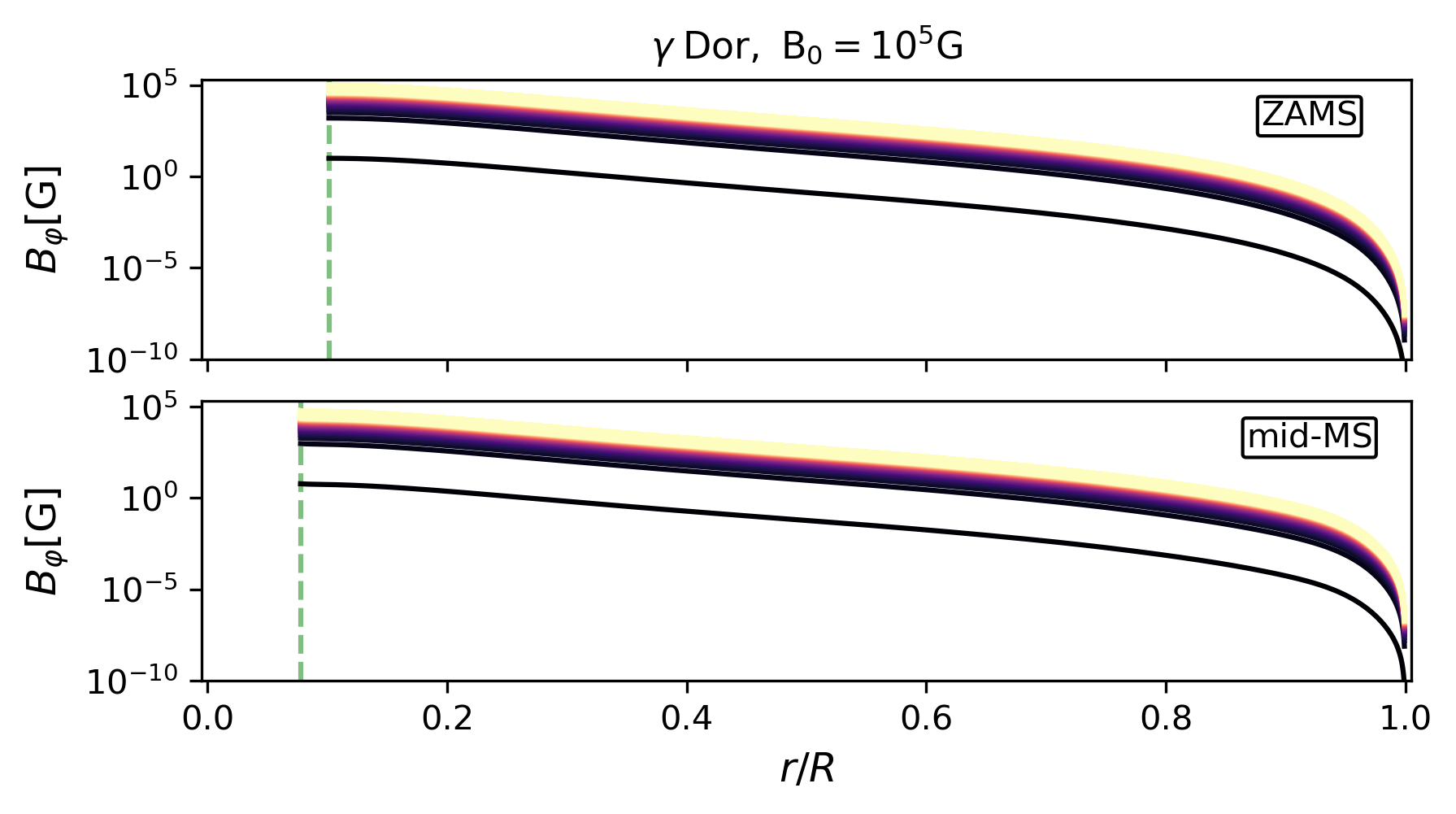}}
    \caption{Fossil toroidal magnetic field profile at $B_0=10^5\,\rm G$ in the radiative zone as a function of the normalised radius at different colatitudes from near the pole (black) to the equator (yellow) for the $1.6{\rm M}_\odot$ $\gamma$\,Dor star model at ZAMS (top) and mid-MS (bottom). The dashed green line indicates the radiative-convective interface.}
    \label{fig:b_phi_dor}
\end{figure}

\begin{figure}
    \centering
     \resizebox{\hsize}{!}{\includegraphics{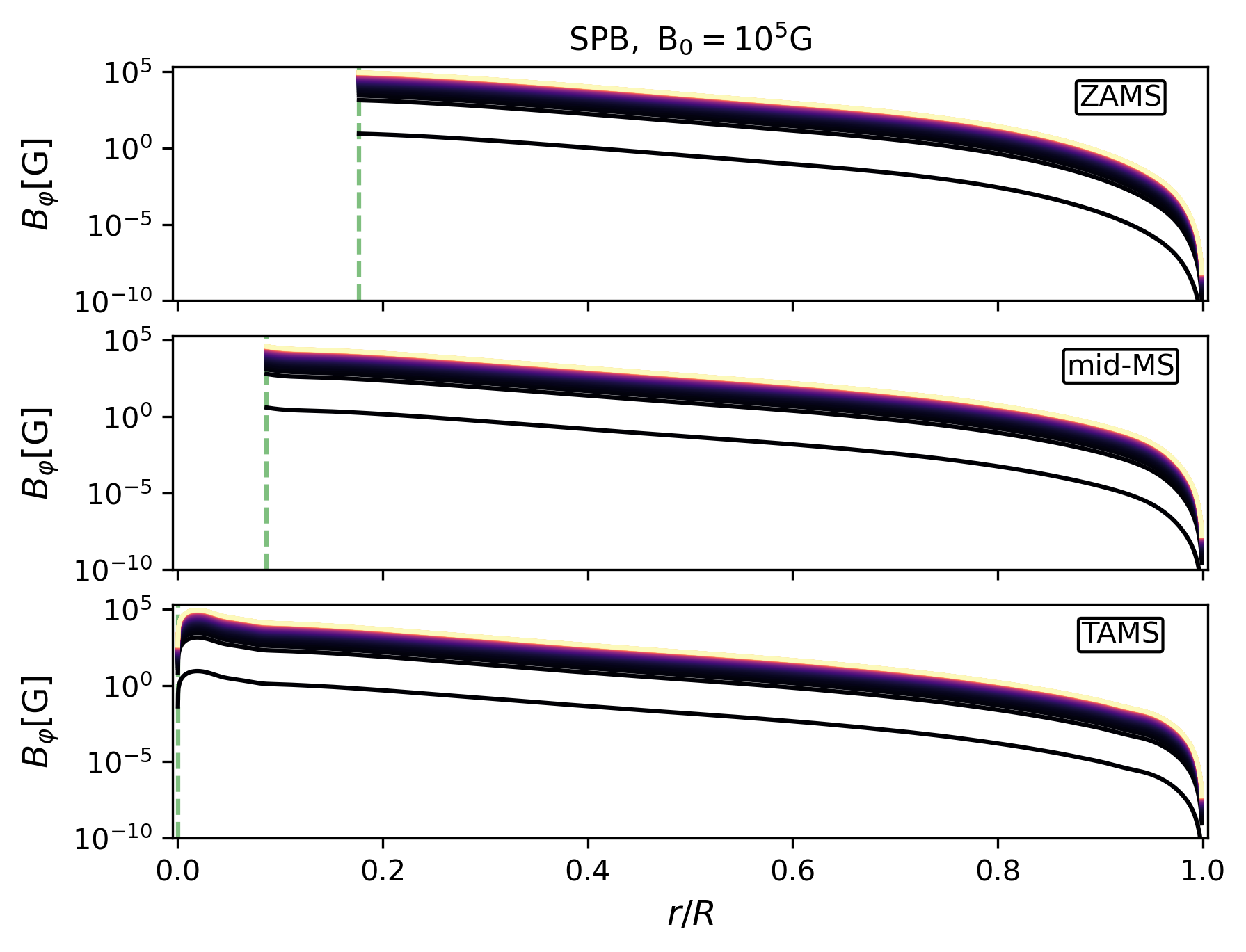}}
    \caption{Same as Fig.\;\ref{fig:b_phi_dor}, but for the $5{\rm M}_\odot$ SPB model at ZAMS (top), mid-MS (middle), and TAMS (bottom).}
    \label{fig:b_phi_spb}
\end{figure}

In Figs.\;\ref{fig:b_phi_dor}\;and\;\ref{fig:b_phi_spb}, we represent the toroidal magnetic field along the evolution of $\gamma$\,Dor and SPB stars with a chosen amplitude $B_0=10^5\,\rm G$. We can see that the magnetic field is the strongest in the inner region of the radiative zone, whereas near the surface it becomes very weak.
Using Eqs.\;(\ref{eq:B_T})\;and\;(\ref{eq:B_duez}), we derive the Alfvén frequency, which becomes independent of $\theta$,
\begin{equation}
    \omega_{\rm A} (r) = \frac{\tilde{b}_\varphi}{\sqrt{\mu_0 \rho_0} r},
\end{equation}
because of the $\sin{\theta}$ dependence of the toroidal field. Here, we assume  that the magnetic permeability of the medium is equal to the one in vacuum.

\subsection{Equations to solve: Uniform rotation case} \label{subsec:equations_to_solve}
Based on seismic constraints, the internal rotation of most intermediate-mass MS stars highlights a weak core-to-surface rotation contrast. Therefore, in this case, we can assume that the rotation is (quasi-)rigid in their radiative zone \citep{Kurtz2014, Saio2015, Murphy2016, Aerts2017, aerts2019}. Such a (quasi-)rigid rotation state is coherent with the rotation profiles resulting from the action of Maxwell stresses of unstable or dynamo-regenerated toroidal fields \citep{MaederMeynet2003,HegerSpruit2005, Eggenberger2005, Fulleretal2019, Petitdemange2021}. In addition, given the topology of the magnetic field that we consider here, the Alfvén frequency is independent of the latitude (see Sect.\;\ref{sec:field_topo}).
From now on, we specifically focus on the case where the rotation is uniform and the Alfvén frequency only depends on the radius.  
Thus, the MLTO simplifies to
\begin{multline}\label{eq:mlto_specific}
       \mathcal{L}^{\rm{magn.}}_{\nu m} = \frac{\omega^2}{\mathcal{A}}\partial_{x}\left[ \frac{1-x^2}{1-\nu_{\rm M}^2 x^2}\partial_{x} \right] 
   +\\ \frac{\omega^2}{\mathcal{A}}\left[m\nu_{\rm M}\frac{1+\nu_{\rm M}^2 x^2}{\left(1-\nu_{\rm M}^2 x^2\right)^2} - \frac{m^2}{\left(1-x^2\right)\left(1-\nu_{\rm M}^2 x^2\right)}\right].
\end{multline}
In the non-magnetic case, this operator reduces to the standard Laplace tidal operator derived by \cite{lee+saio1997}
\begin{multline}\label{eq:slte}
    \mathcal{L}^{\rm{stand.}}_{\nu m}  = \partial_{x}\left[ \frac{1-x^2}{1-\nu^2 x^2}\partial_{x} \right] +
     m\nu\frac{1+\nu^2 x^2}{\left(1-\nu^2 x^2\right)^2} -\\ \frac{m^2}{\left(1-x^2\right)\left(1-\nu^2 x^2\right)}.
\end{multline}
In the general case and specifically because of the differential rotation, we indexed the operator and the variables by the wave frequency in the inertial frame $\omega^{\mathrm{in}}$ because it is the only constant frequency in that case. However, in the uniformly rotating case, it is more convenient to use the spin parameter $\nu$ for the indexation. 
In this specific case, we can derive the analytical expression of the asymptotic frequencies in the rotating frame. Hence, from Eq.\;(\ref{eq:quantisation}) we recover
\begin{equation}\label{eq:frequencies}
    \omega_{n k m}=\frac{1}{(n+1 / 2) \pi} \bigintssss_{r_{1}}^{r_{2}} \frac{N\left(r\right)}{r}\sqrt{\Lambda_{\nu_{n} k m}^{\rm{magn.}}\left(r\right)}\;\mathrm{d} r,
\end{equation}
with
\begin{equation}
    \nu_n=\nu_{nkm}=\frac{2 \Omega}{\omega_{nkm}}.
\end{equation}
We obtain here the same expression as in the non-magnetic case. The only difference is that the eigenvalues are modified due to the magnetic field so they are no longer constant but depend on the radius. 
This expression is implicit because $\Lambda_{\nu_{n} k m}^{\rm{magn.}}$ depends on $\nu_{n}$, which in turn depends on $\omega_{n k m}$. Therefore, Eq.\;(\ref{eq:frequencies}) must be solved numerically as explained in Sect.\;\ref{subsec:period_spacing}.

\subsection{Frequency hierarchy}
\begin{figure*}
    \centering
     \includegraphics[width=13cm]{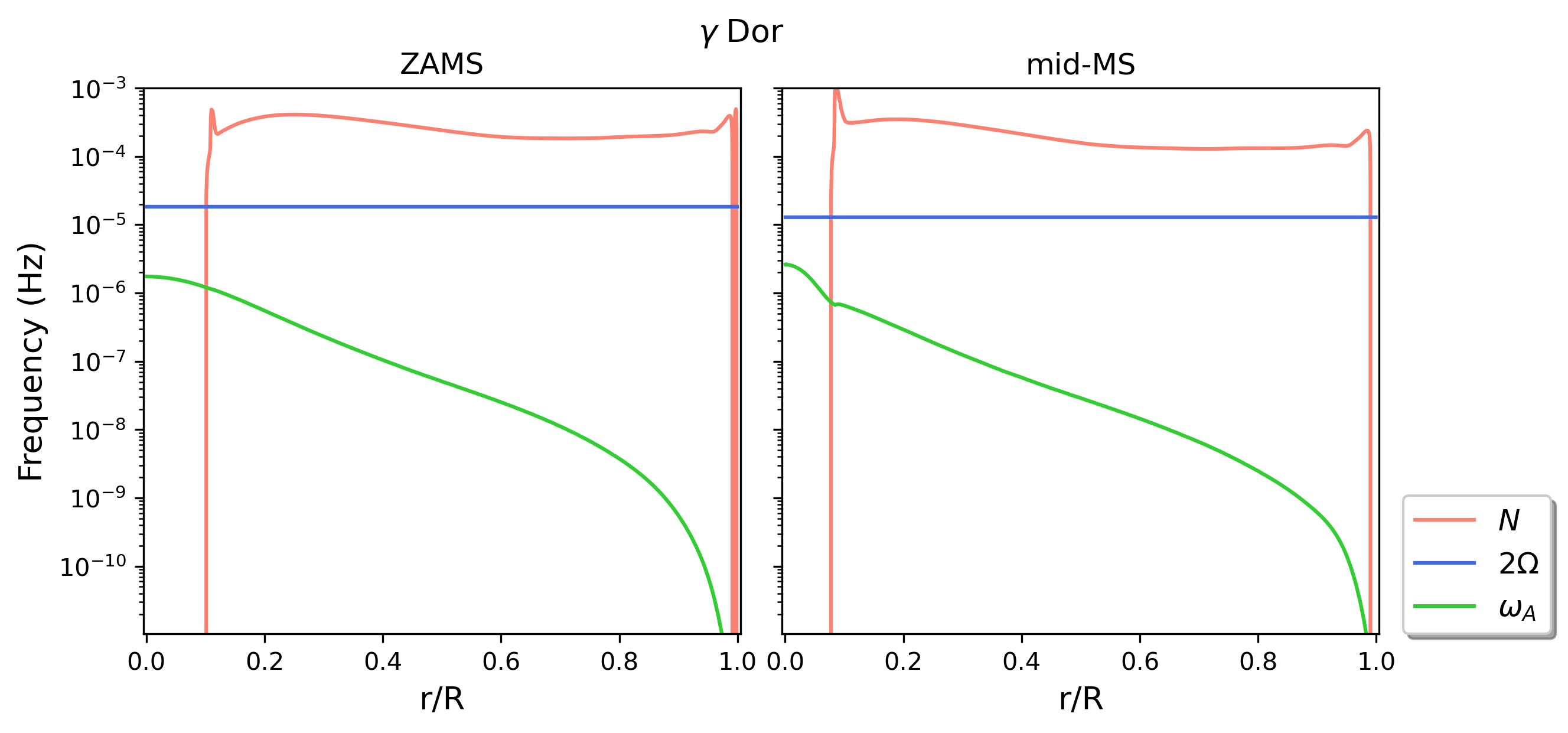}
      \includegraphics[width=18.4cm]{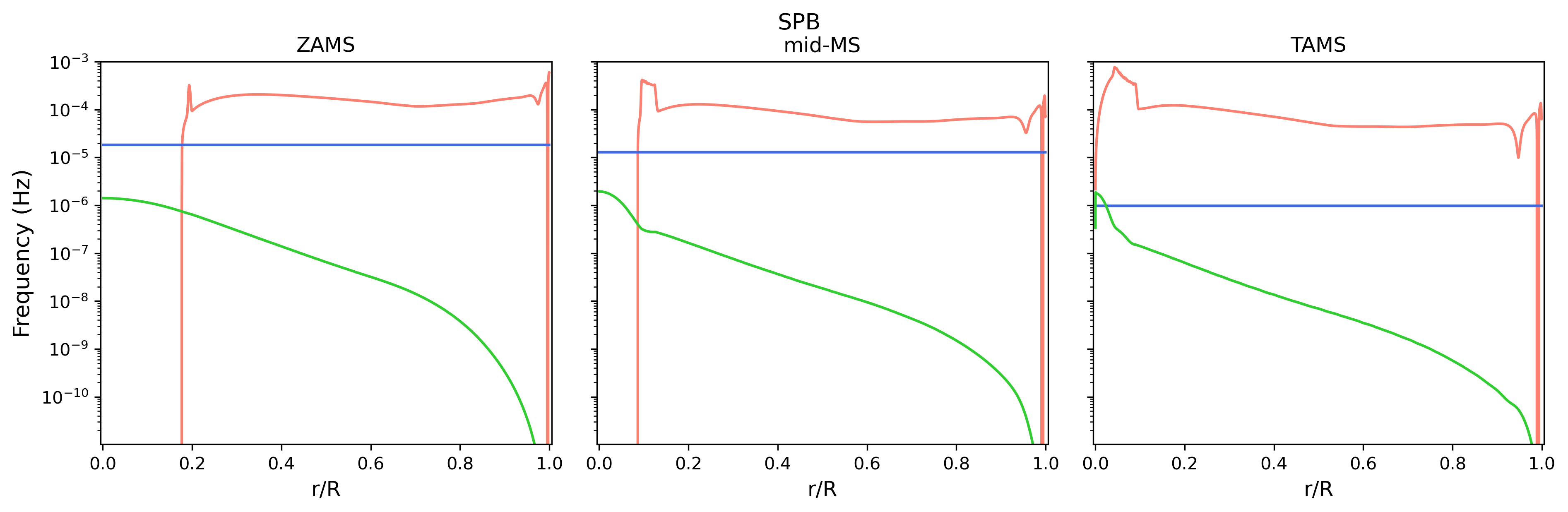}
    \caption{Profiles of the Brunt–Väisälä ($N$), inertial ($2\Omega$), and Alfvén ($\omega_{\rm A}$) frequencies  for the  $1.6{\rm M}_\odot$ $\gamma$\,Dor (top) and $5{\rm M}_\odot$ SPB (bottom) models at different evolutionary stages rotating at the rotation rates specified in Table \ref{tab:rot_tar} for $B_0=10^5\,\rm G$.}
    \label{fig:freq_dor}
\end{figure*}

In order to ensure that we work within the validity domain of the magnetic TAR, a set of frequency hierarchies must be verified. The most constraining one is $N\gg 2\Omega$. In this perspective, we determine the maximal rotation rate that 
complies with $\min\left(N/2\Omega_{\rm max}^{\rm TAR}\right)=\beta$, with $\beta$ a chosen threshold 
such that the stratification dominates over the rotation. We found that the age of the star and the maximal allowed rotation rate are inversely proportional.
We summarise the values of $\Omega_{\rm max}^{\rm TAR}$ for the $\gamma$\,Dor and SPB models at different life phases and for $\beta=10$ in Table\;\ref{tab:rot_tar}.
\begin{table}
    \caption{Maximal values of the rotation rate in which the (magnetic) TAR is applicable, $\Omega_{\rm max}^{\rm TAR}$.}
    \label{tab:rot_tar}
    \centering
    \begin{tabular}{l | c c}
        \hline \hline
        \diagbox{Phase}{Star} & $\gamma$\,Dor & SPB \\
        \hline
        near-ZAMS & $22.19\Omega_{\odot}$  &$11.38\Omega_{\odot}$ \\
        mid-MS & $15.56\Omega_{\odot}$ & $3.94\Omega_{\odot}$ \\
        near-TAMS & \ldots & $1.18\Omega_{\odot}$\\
        \hline
    \end{tabular}

    \tablefoot{$\Omega_{\odot}\approx\unit{4.14\times 10^{-7}}{\hertz}$ denotes the solar rotation.}
\end{table}
The spherical background assumption is fulfilled since $\Omega/\Omega_{\mathrm{K}}< 3\%$. 
In Fig.\;\ref{fig:freq_dor}, we show the characteristic frequency profiles in the case where these criteria are met. As is well known, the radiative envelope gets bigger and the convective core shrinks considerably as the star evolves. We find that the Alfvén frequency is always lower or has the same order of magnitude as the characteristic frequency of rotation $2\Omega$ when the amplitude $B_0$ is smaller or around $10^5\;\rm G$. 
This ensures that the other hierarchies are also met if the low-frequency hierarchy ($\omega\ll N$) is fulfilled, which we check a posteriori.

\subsection{Eigenvalues and Hough functions}\label{sect:hough_eigen}
\begin{figure}
    \centering
     \resizebox{\hsize}{!}{\includegraphics{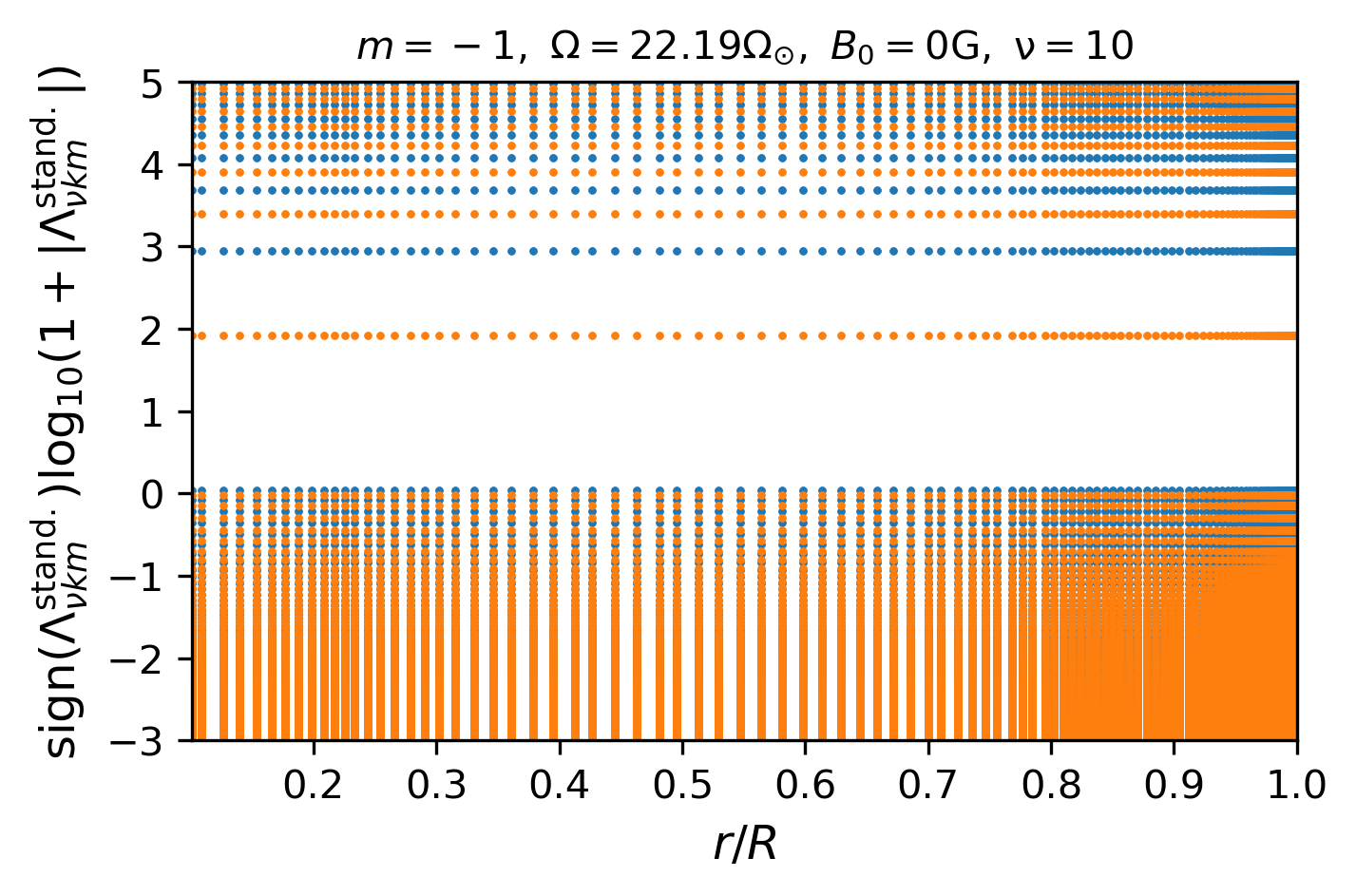}}
     \resizebox{\hsize}{!}{\includegraphics{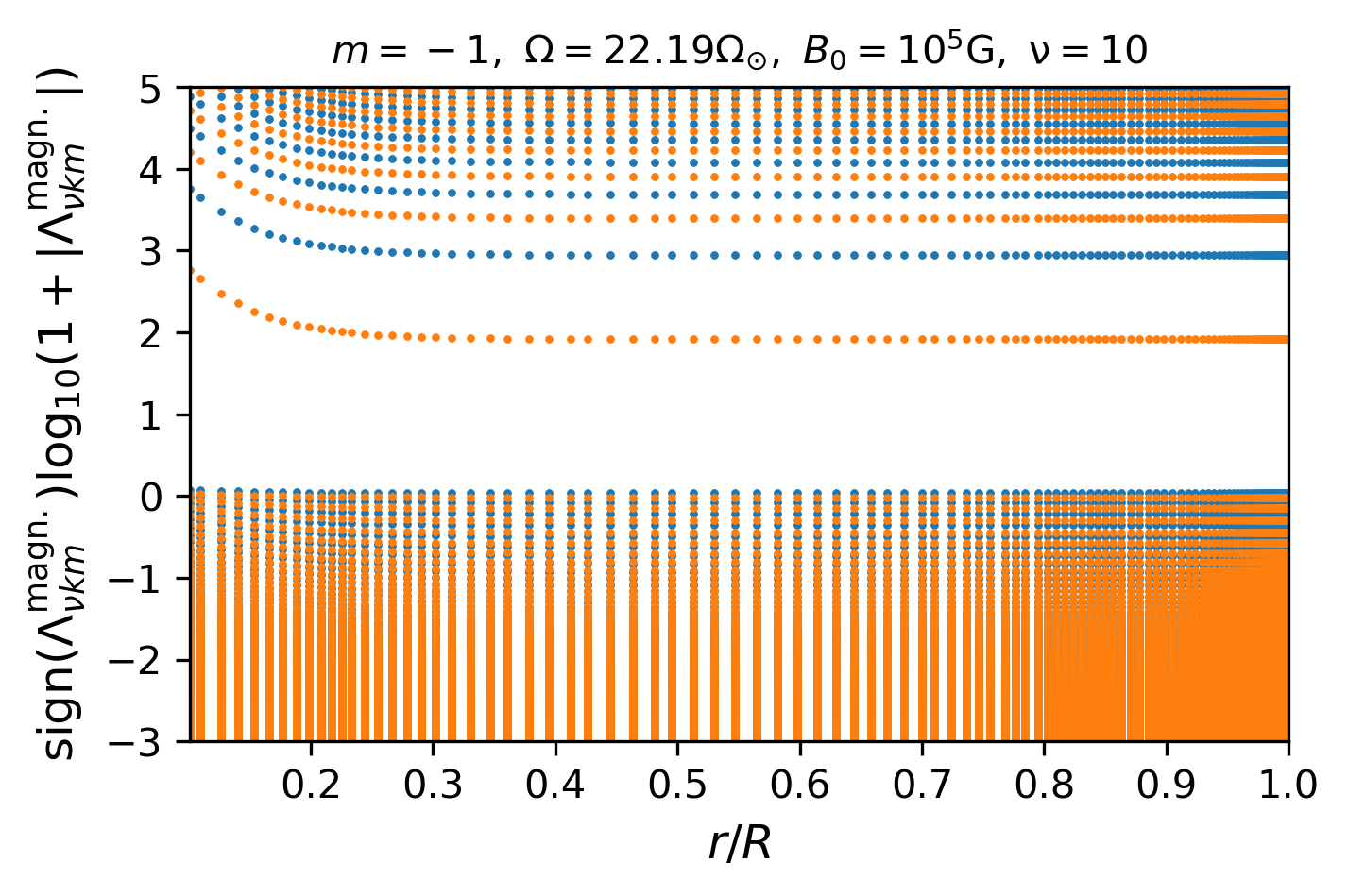}}
    \caption{Spectrum of the Laplace tidal equation in the non-magnetic (top) and magnetic (bottom) cases as a function of the radius for $m=-1$, $\nu=10,$ and $ \Omega = 22.19\,\Omega_{\odot} $ using the $1.6{\rm M}_\odot$ $\gamma\,$Dor model at ZAMS. Blue (respectively, orange) dots correspond to modes with even (respectively, odd) Hough functions.}
    \label{fig:eigenvalues}
\end{figure}
\begin{figure}
    \centering
     \resizebox{\hsize}{!}{\includegraphics{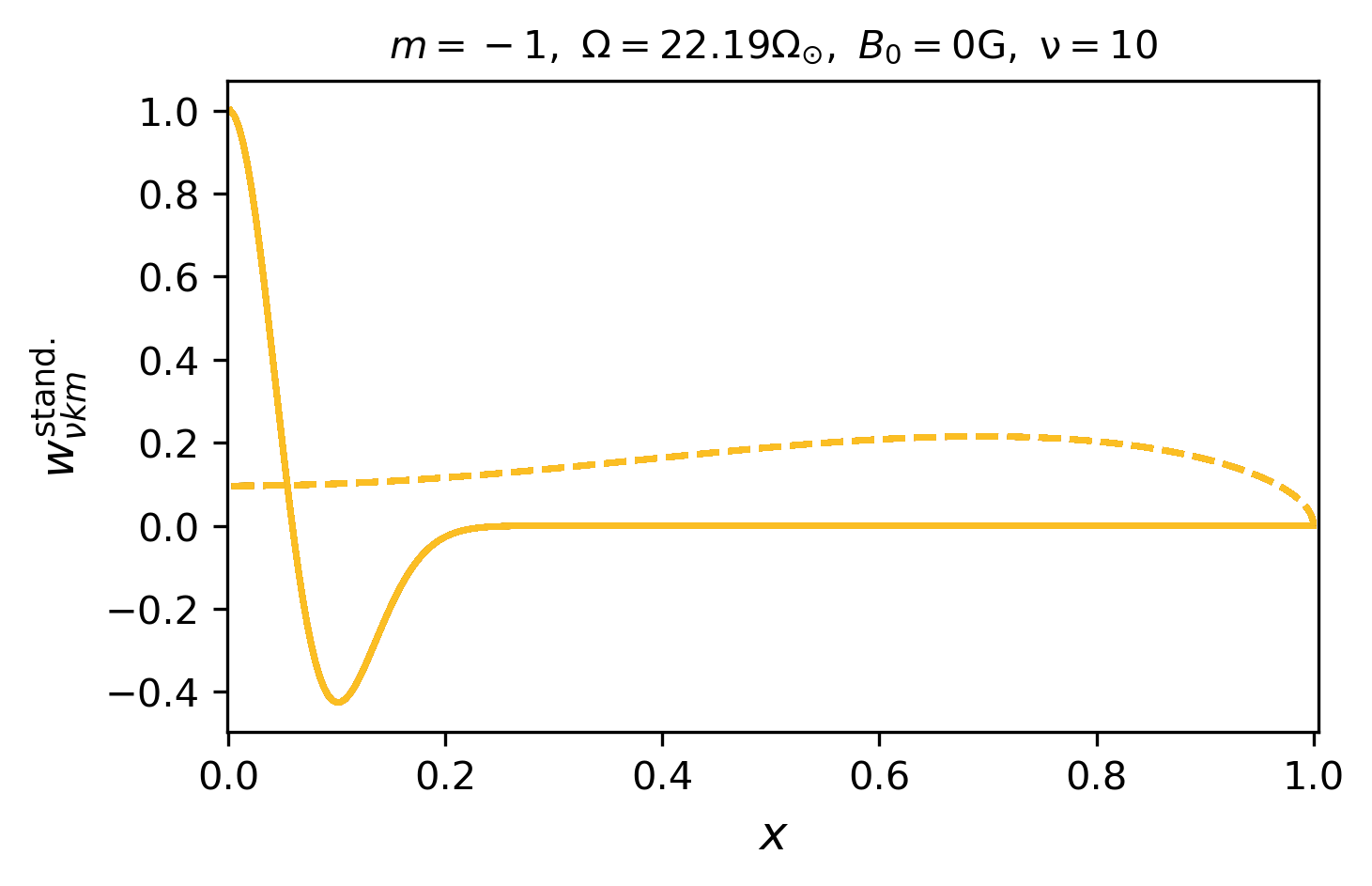}}
     \resizebox{\hsize}{!}{\includegraphics{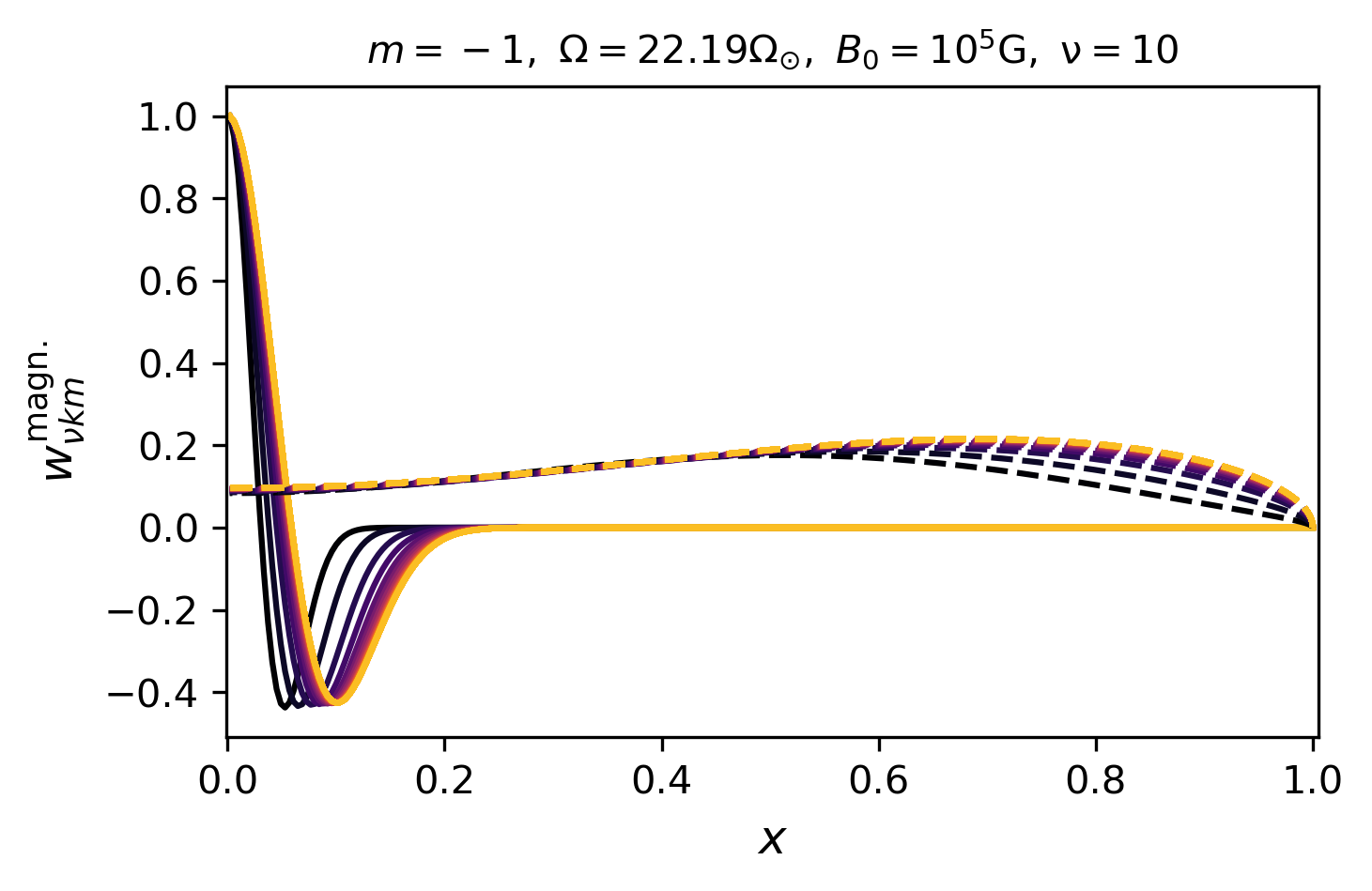}}
    \caption{Hough functions (normalised to unity) in the non-magnetic (top) and magnetic (bottom) cases as a function of the horizontal coordinate $x$ at different radii from the radiative-convective interface (dark blue) to the surface (orange) for $m=-1$, $\nu=10,$ and $ \Omega = 22.19\,\Omega_{\odot} $ at $B_0=10^5\,\rm G$ using the $1.6{\rm M}_\odot$ $\gamma\,$Dor model at ZAMS. 
    Solid lines correspond to gravity-like solutions  $(k = 0), $ and dotted lines correspond to Rossby-like solutions $(k = -2)$.}
    \label{fig:hough}
\end{figure}
The MLTE is an eigenvalue problem. To solve it numerically, we use the Chebyshev collocation method, which was first applied to solve the standard Laplace tidal equation by \cite{Boyd1976}. As in \cite{mathis+prat2019}, \cite{Henneco2021}, and \cite{dhouib2021a,dhouib2021b}, we use an implementation similar to the one used by \cite{wang2016}. 
We solve the MLTE for different radii $r$, spin parameters $\nu$, rotation rates $\Omega$ and magnetic field amplitude scaling factors $B_0$. We focus on the impact of 
the toroidal magnetic field on the eigenvalues and Hough functions. As a representative example, we use the $1.6{\rm M}_\odot$ $\gamma$\,Dor model at the  ZAMS to show 
these quantities
in Figs.\;\ref{fig:eigenvalues}\;and\;\ref{fig:hough}. We obtain similar results with the other models. In these figures, we can see the influence of the magnetic field when $B_0=10^5\,\rm G$. 
Figure\;\ref{fig:eigenvalues} reveals that the eigenvalues are modified by the magnetic field, which introduces a strong radial dependence in the inner region of the star (close to the interface between the radiative envelope  and the convective core). Close to the surface, the influence of the magnetic field on the eigenvalues is very weak, almost nonexistent. This is expected since the toroidal magnetic field is weak far from the radiative-convection interface because $\tilde{b}_\varphi\left(r\right) \to 0$ when $r \to R$ (Figs.\;\ref{fig:b_phi_dor}\;and\;\ref{fig:b_phi_spb}).

We use the convention $m\nu>0$ (resp. $m\nu<0$) for prograde (resp. retrograde) modes.
We can distinguish mainly two types of solutions. The first ones are gravity-like solutions ($k\geq0$), which exist in the non-rotating non-magnetic case. They correspond to internal gravity modes ($\rm g$\,modes) modified by the Coriolis acceleration and the Lorentz force (MGI modes). The second type of solutions are Rossby-like ones ($k<0$), which exist only in the rotating case when $|\nu|>1$. If they are retrograde and have positive eigenvalues, these solutions correspond to Rossby (quasi-inertial) modes ($\rm r$\,modes) modified by the Lorentz force and the buoyancy \citep{saio2018}. The third family of solutions are those with negative eigenvalues. They correspond to evanescent modes or rotationally stabilised convective modes. These modes are out of the scope of this work since we focus here on the stably stratified regions only. We refer the reader to \cite{lee2019}, \cite{lee+saio2020}, and references therein for detailed discussions of this third family of solutions.

In Fig.\;\ref{fig:hough}, we show the magnetic Hough functions for gravity like-solutions (retrograde sectoral $\rm g$\,mode with $\{k = 0, m=-1\}$) and Rossby like-solutions ($\rm r$\,mode with $\{k = -2, m=-1\}$). The yellow curve represents the Hough functions at the surface of the star, which is nearly the same as in the non-magnetic case due to the weak surface magnetic field. The behaviour of these eigenfunctions changes the most in the inner regions (blue curve). More specifically, the magnetic Hough functions migrate inwards, towards the equator ($x=0$), causing a narrowing of their overall shape. Furthermore, the gravity like-solutions are trapped near the equatorial plane in the regime ($\nu_{\rm M}>1$). In fact, regions where $D_{\rm M}\ge0$ correspond to regions where the gravity like-solutions are propagative, whereas they become evanescent where $D_{\rm M}<0$. Therefore, when $\nu_{\rm M}>1$, the gravity-like solutions become trapped in an equatorial belt \citep{mathis2009,Mathis&deBrye2011}. In contrast to gravity-like solutions, Rossby-like solutions are found in the  whole spherical domain and they have in general no tendency to be trapped in the equator but rather become concentrated towards mid-latitudes. The Rossby-like solution $\{k = -1, m=-1\}$, however, is trapped at equator due to its large eigenvalue \citep{saio2018,Saio2018b}.
Under the effect of the magnetic field, the trapping of the $\rm g$\,modes becomes more efficient as the critical reduced colatitude  $x_{\rm c}=\cos\theta_{\rm c}=\left|\nu_{\rm M}\right|^{-1}$ \citep{Mathis&deBrye2011}, which corresponds to the turning point of $D_{\rm M}$,  decreases. Thus, as the magnetic spin parameter $\nu_{\rm M}$ increases, these modes tend to be more confined around the equator.

\subsection{Asymptotic period spacing pattern}\label{subsec:period_spacing}
\begin{figure}
    \centering
     \resizebox{\hsize}{!}{\includegraphics{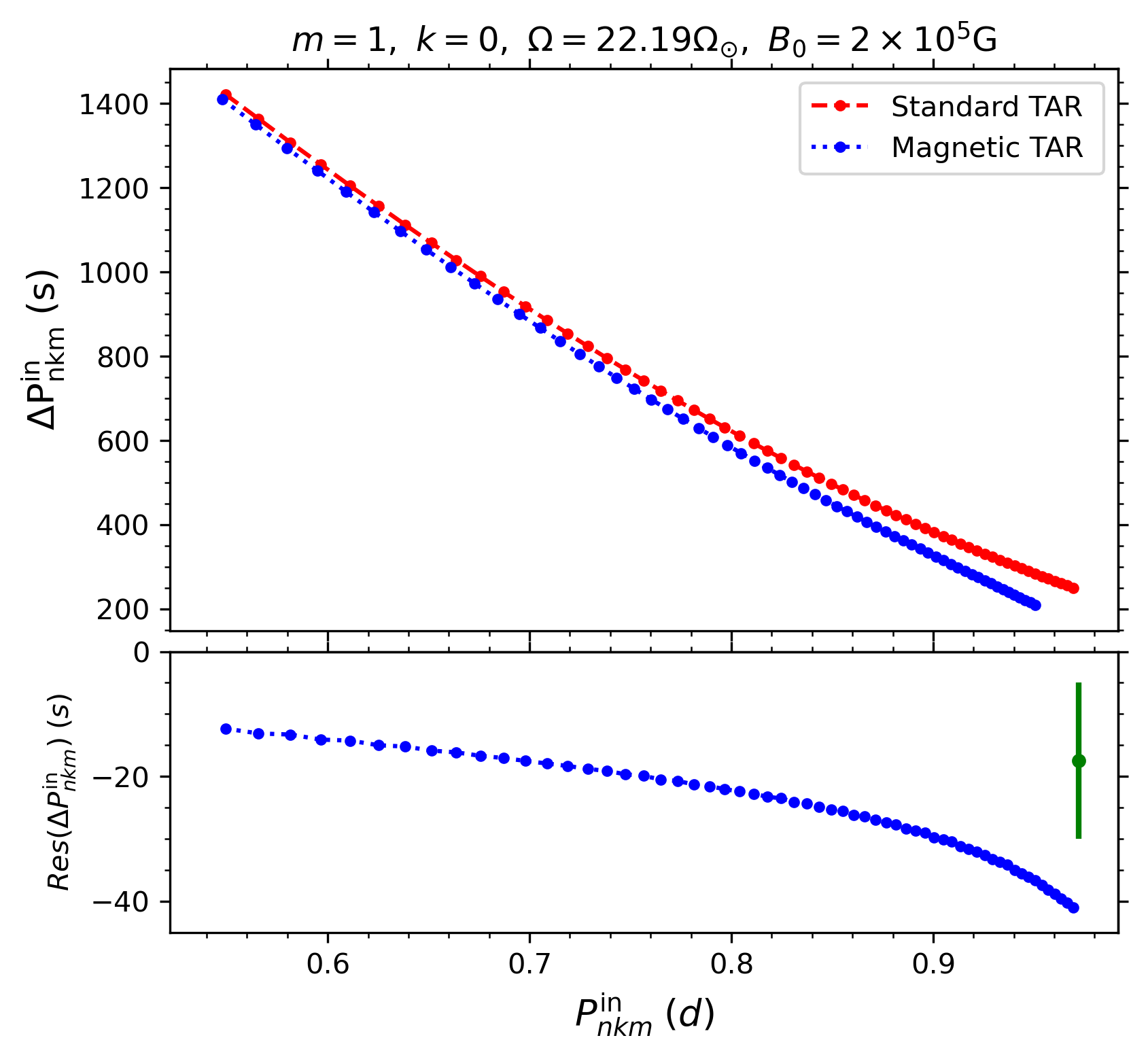}}
    \caption{Period spacing pattern in the inertial frame for $\rm g$ modes $\{k=0,m=1\}$ computed using the standard TAR (red line) and the magnetic TAR (blue line) with the $1.6{\rm M}_\odot$ $\gamma\,$Dor model at ZAMS. The bottom panel shows the differences with respect to the standard TAR period spacing  pattern. The vertical green bar represents an average error of measured period spacings for a sample of 40 $\gamma\,$Dor stars \citep{vanreeth2015}.}
    \label{fig:delta_p_m1}
\end{figure}
\begin{figure}
    \centering
     \resizebox{\hsize}{!}{\includegraphics{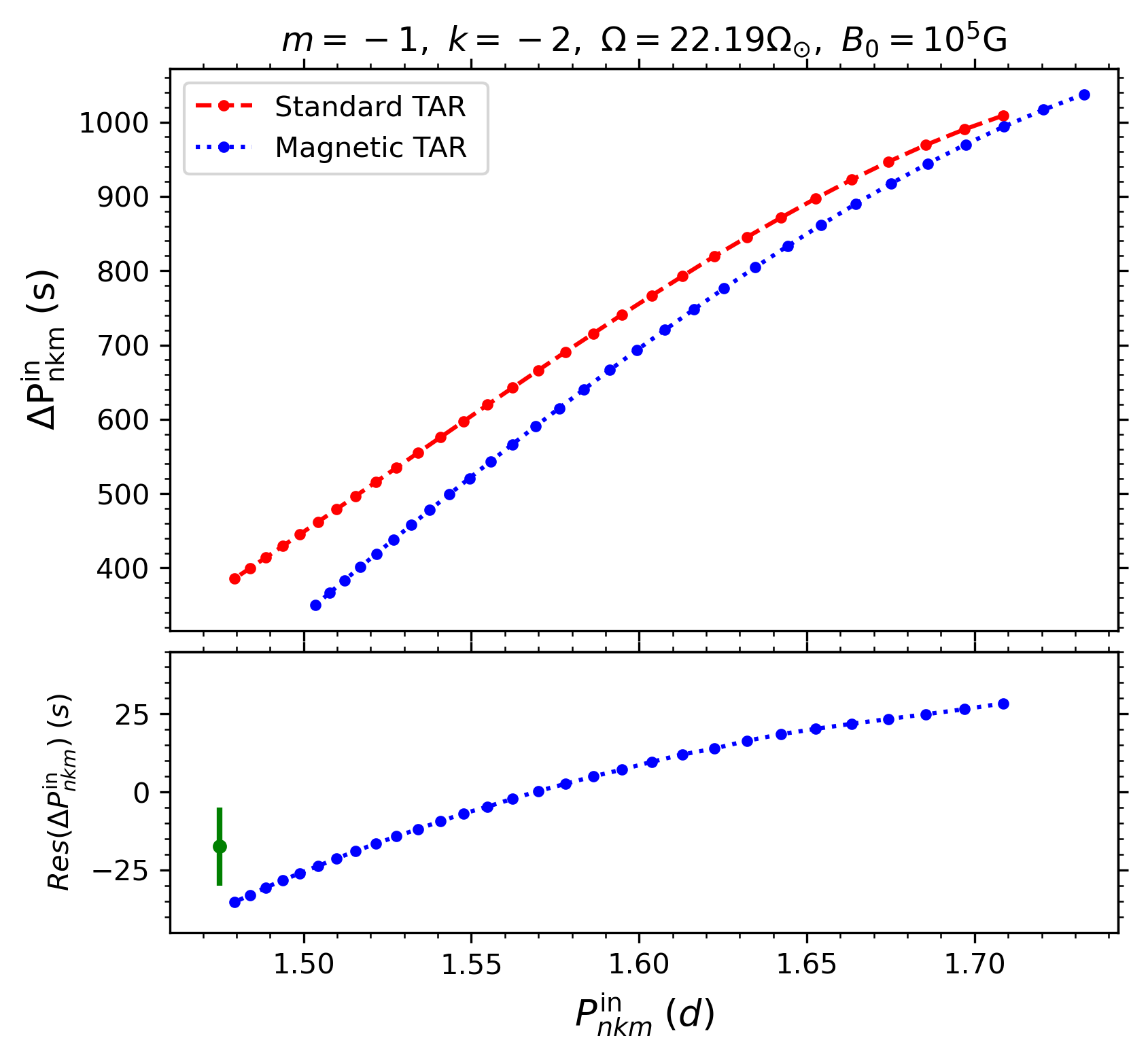}}
    \caption{Same as Fig.\;\ref{fig:delta_p_m1}, but for $\rm r$ modes $\{k=-2,m=-1\}$.}
    \label{fig:delta_p_m-1}
\end{figure}

The goal of this section is to assess the asymptotic period spacing of the modes as a function of their period,  which is a standard tool used to interpret oscillations spectra, of $\rm g$ and $\rm r$ modes \citep{Miglio2008,Bouabidetal2013,vanreeth2015,aerts2021}. The period spacing is defined as the period differences between modes of consecutive radial orders $n$ with the same azimuthal order  $m$ and latitudinal index $k$ (linked to the latitudinal order $\ell=\left|m\right|+k$ \citep{lee+saio1997}, which is used in the non-rotating non-magnetic case),
\begin{equation}
    \Delta P _{nkm}= P_{n+1km}- P_{nkm}.  
\end{equation}
To evaluate this quantity, we first compute the asymptotic frequencies
by solving the Eq.\;(\ref{eq:frequencies}) numerically as we discussed in Sect.\;\ref{subsec:equations_to_solve}. To do so, we use the method developed by \cite{Henneco2021} (more details on this method can be found in their Appendix\;B) as in \cite{dhouib2021a,dhouib2021b}.
We focus our attention on three modes that are most often observed in $\gamma$\,Dor and SPB stars \citep{Papics2017, saio2018, Li2020}: $\rm g$\,modes with $\{k = 0, m=1\}$ and $\{k = 0, m=2\}$, and $\rm r$\,modes with $\{k = -2, m=-1\}$. The modes that we compute here have radial orders between 20 and 80. The minimum order is chosen to make sure that the Cowling approximation is met since this approximation is only valid for high radial orders (see for instance \cite{dhouib2021b}). \cite{Li2020} found that the majority (more than 68\%) of the observed modes  in $\gamma\,$Dor stars have radial orders between 20 and 70 (their Fig.\;17).

Figures\;\ref{fig:delta_p_m1} and\;\ref{fig:delta_p_m-1} show the magnetic period spacing pattern with their standard non-magnetic counterpart. We find that, for $\rm g$ modes, the spacing values decrease under the influence of the toroidal magnetic field. This decrease is largest at the highest radial orders (longest mode periods) for which the action of the Lorentz force is strongest. For the $r$ modes, the spacing values decrease for short periods (high radial orders) and increase for long periods (low radial orders).
We also find that the toroidal magnetic field does not have a distinctive signature on the period spacing pattern, rather that it introduces a shift. Such an effect can also be introduced by other physical effects, such as the centrifugal acceleration or differential rotation. Using a perturbative approach,
\cite{Prat2019, Prat2020} and \cite{VanBeeck2020} found that the signature of a fossil poloidal–toroidal magnetic field in the period spacing morphology of $\rm g$\;modes differs appreciably from the signature due to rotation. In addition to the frequency shifts that increase with increasing the radial order, they found characteristic magnetic sawtooth-like features for high-radial-order modes. We provide a few plausible explanations for the absence of such features in the period spacings for the few stellar models considered in this work. The features may be the signature of the poloidal component of the magnetic field. On the other hand, the features may have been introduced by the limitation of the perturbative approach. Finally, and most plausibly, the difference in saw-tooth features could be related to dips in the period spacing patterns due to mode trapping caused by chemical gradients left behind by the shrinking convective core, as predicted theoretically by \citet{Miglio2008} and \citet{Bouabidetal2013}, found numerically by \citet{Pedersen2018} and \citet{Michielsen2019}, and observed in many $\gamma\,$Dor \citep{Mombarg2021} and SPB \citep{Pedersen2021} stars. Future dedicated
analyses are planned to unravel these hypotheses as follow-up studies of the current theoretical work.

\subsection{Detectability of the magnetic signature}
\begin{figure}
    \centering
     \resizebox{\hsize}{!}{\includegraphics{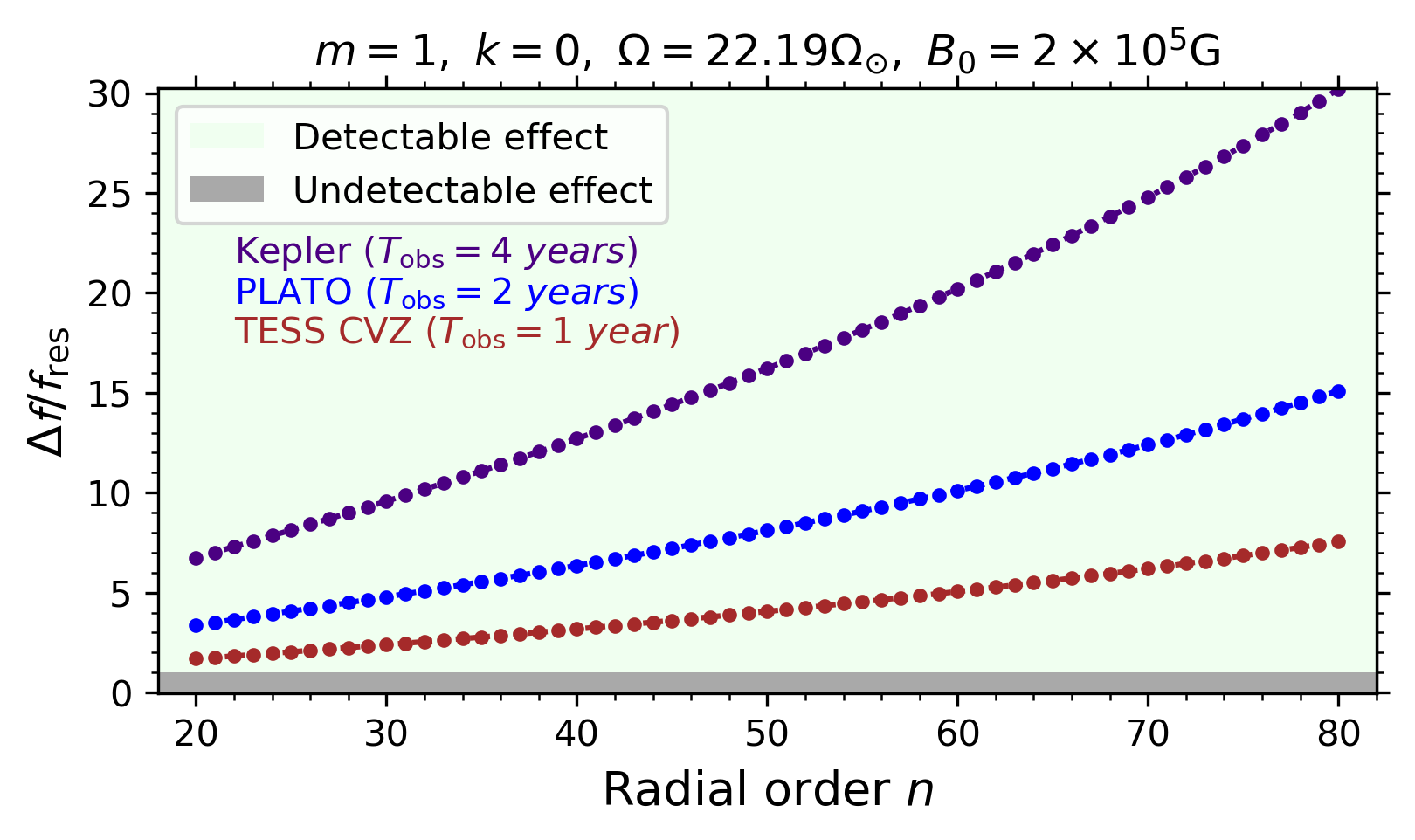}}
      \resizebox{\hsize}{!}{\includegraphics{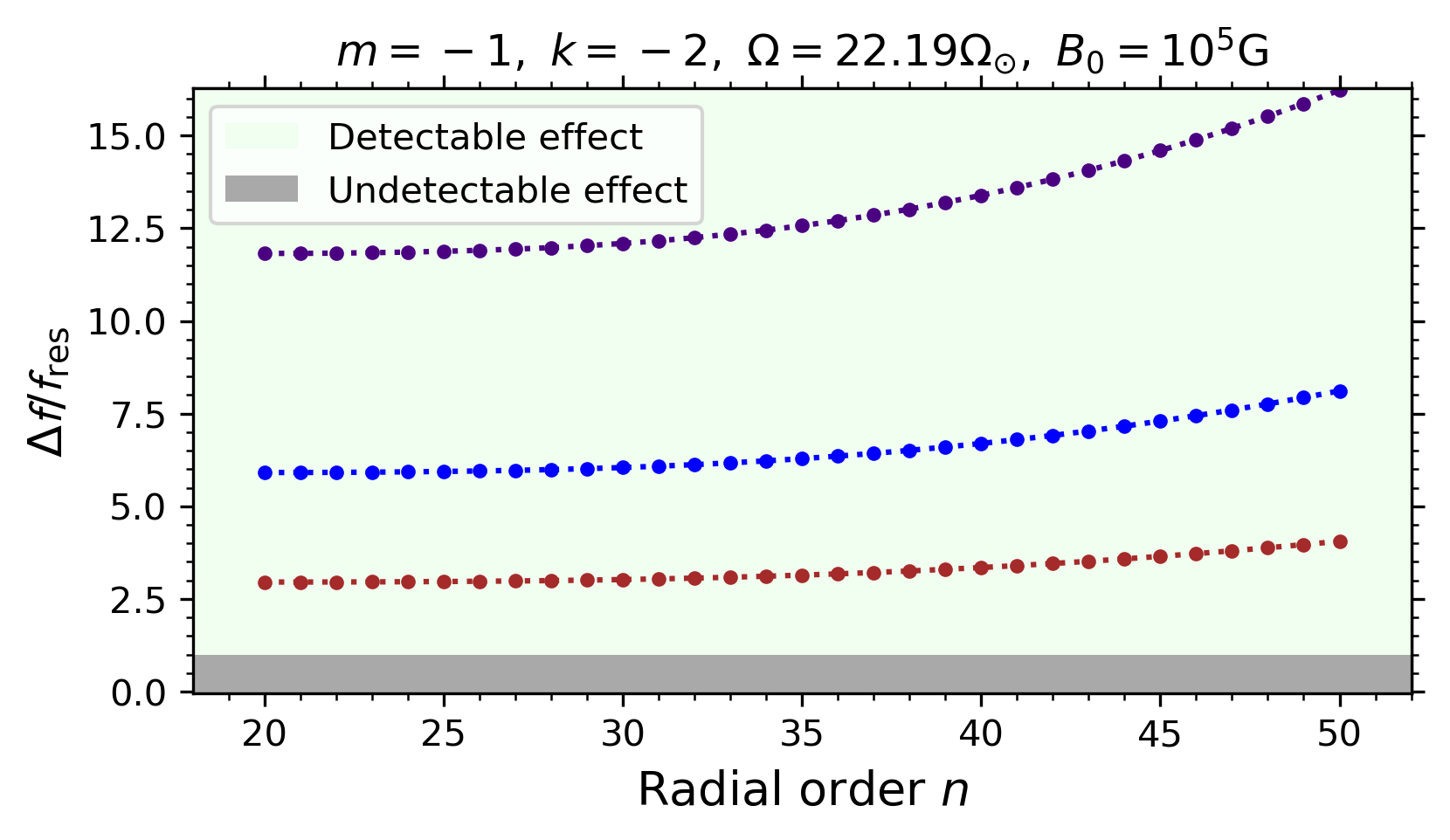}}
    \caption{Detectability of the toroidal magnetic field effect on the $\rm g$ modes $\{k=0,m=1\}$ (top) and the $\rm r$ modes $\{k=-2,m=-1\}$ (bottom) as a function of the radial order  $n$ using the $1.6{\rm M}_\odot$ $\gamma\,$Dor model at the ZAMS based on the  frequency resolution of  \textit{Kepler} (indigo line), PLATO (blue line), and TESS CVZ (brown line).}
    \label{fig:detect}
\end{figure}
To quantify the theoretical detectability of the toroidal magnetic field, we compare the frequency differences between asymptotic frequencies calculated with the standard TAR and those calculated with the magnetic TAR for each radial order,
\begin{equation}
    \Delta f(n) = |f_{\rm magn.}(n) - f_{\rm stand.}(n) |,
\end{equation}
to the one-year Transiting Exoplanet Survey Satellite Continuous Viewing Zone \citep[TESS CVZ;][]{ricker2015}, the two-year PLAnetary Transits and Oscillations of stars \citep[PLATO;][]{Rauer2014,Rauer2016}, and the four-year \textit{Kepler} \citep{borucki2010, Howell2014} frequency resolutions $f_{\mathrm{res}}=1/T_{\mathrm{obs}}$, with $T_{\mathrm{obs}}$ the observation time.
In this way, we are able to deduce the range in radial orders $n_{\mathrm{detect,min}}$ for which the frequency differences are expected to be detectable:
\begin{equation}
    \Delta f(n)>f_{\mathrm{res}}.
\end{equation}
We show these results in Fig.\;\ref{fig:detect}
for $\rm g$ $\{k=0,m=1\}$ and $\rm r$ $\{k=-2,m=-1\}$ modes using the $1.6{\rm M}_\odot$ $\gamma$\,Dor model with $B_0=2\times 10^5\;\rm G$ located at the  ZAMS.
We can see that in this case the magnetic effect is in principle detectable for all radial orders using nominal TESS CVZ, \textit{Kepler} and PLATO light curves. The signature of the magnetic effect increases as the radial order of the mode increases. To get a consistent idea on the magnetic signature, we explore a wide range of parameters within the validity domain of the magnetic TAR using our equilibrium models. This exploration is summarised in Tables\;\ref{table:detect_gamma} and\;\ref{table:detect_spb}. 
We find that magnetic fields with amplitudes up to about $10^4\,\rm G$ have no detectable signature on  $\rm g$\ and $\rm r$ modes. A most important conclusion of our work is that, for all stronger magnetic fields, its signatures are predicted to be detectable in observations.

High values of $B_0$ can produce a magnetic critical layer for which $\mathcal{A}=0$, which limits the formation of the modes. The $\rm r$ modes undergo this resonance for lower $B_0$ values than the $\rm g$ modes ($B_0 \sim 5\times10^5\,\rm G$ for $\rm g$ modes while for $\rm r$ modes $B_0 \sim 10^5\,\rm G$).
Nevertheless, Rossby modes can still be detectable for some radial orders at these amplitudes.
For low rotation rates ($\Omega/\Omega_{\odot}=1.18$), no resonant $\rm r$\,modes were found. This translates into the opportunity to deduce strong constraints on the free parameters of the field and the rotation to avoid a magnetic critical layer. In particular, we find that an increase in the amplitude of the magnetic field limits the propagation of some modes because of the presence of the resonance ($n_{\rm max}$ decreases with increasing $B_0$). In addition, increasing the rotation rate of the star allows us to increase the magnetic field amplitude, but the rotation rates listed in Table \ref{tab:rot_tar} cannot be exceeded. We emphasise that the validity of the theory at low frequency decreases for advanced evolutionary stages (TAMS) for some $\rm g$\,modes, where the wave frequency gets closer to the Brunt–Väisälä frequency such that the TAR is no longer reliable. This is potentially caused by the fact that the radiative zone expands, leading to higher values of $N$ as shown in Fig.\;\ref{fig:freq_dor}. In the earlier evolutionary stages, the validity tends to be fulfilled for the high radial orders ($N/\omega>10$) but less so for low radial orders ($n \sim 20$) since ($N/\omega\sim10$) of some $\rm g$\,modes. However, the TAR is still reliable in these cases. We found it to be fulfilled ($N/\omega\gg10$) for $\rm r$\,modes in all the cases.

If strong fossil toroidal fields ($B_0\geq 10^5\rm G $) exist in MS stars, they should readily be detectable from current TESS CVZ and \textit{Kepler} and future  PLATO photometric light curves of $\rm g$-mode pulsators. For $\rm r$ modes, this is even the case for weaker magnetic fields ($B_0\sim 6\times10^4\rm $) since at higher amplitudes we encounter magnetic critical layers ($\omega=\omega_{\rm A}$). The observational challenge is to unravel the magnetic influence from other uncalibrated astrophysical effects occurring in stellar interiors models, such as the presence of rotational mixing and atomic diffusion, which also induce signatures in predicted period spacing values \citep{aerts2021, Mombarg2022}.

The amplitudes of the magnetic fields whose signatures we expect to be detectable using our magnetic TAR formalism are in agreement with the ones found by
\cite{HegerSpruit2005}, \cite{Maeder&Meynet2005}, \cite{Fulleretal2019}, and \cite{Petitdemange2021}. 
These authors found the toroidal magnetic configurations in rotating early-type stars to have strengths between $10^4$ to $10^6\;\rm G$ in their numerical simulations. Therefore, we expect that these fields can be detected and thoroughly investigated using the theoretical magnetic TAR formalism developed in this work, with the potential to unravel the internal magnetism of stars throughout their MS lifetime.

\section{More general magnetic configuration: Hemispheric field}\label{sect:hemispheric}

\begin{figure}
    \centering
     \resizebox{\hsize}{!}{\includegraphics{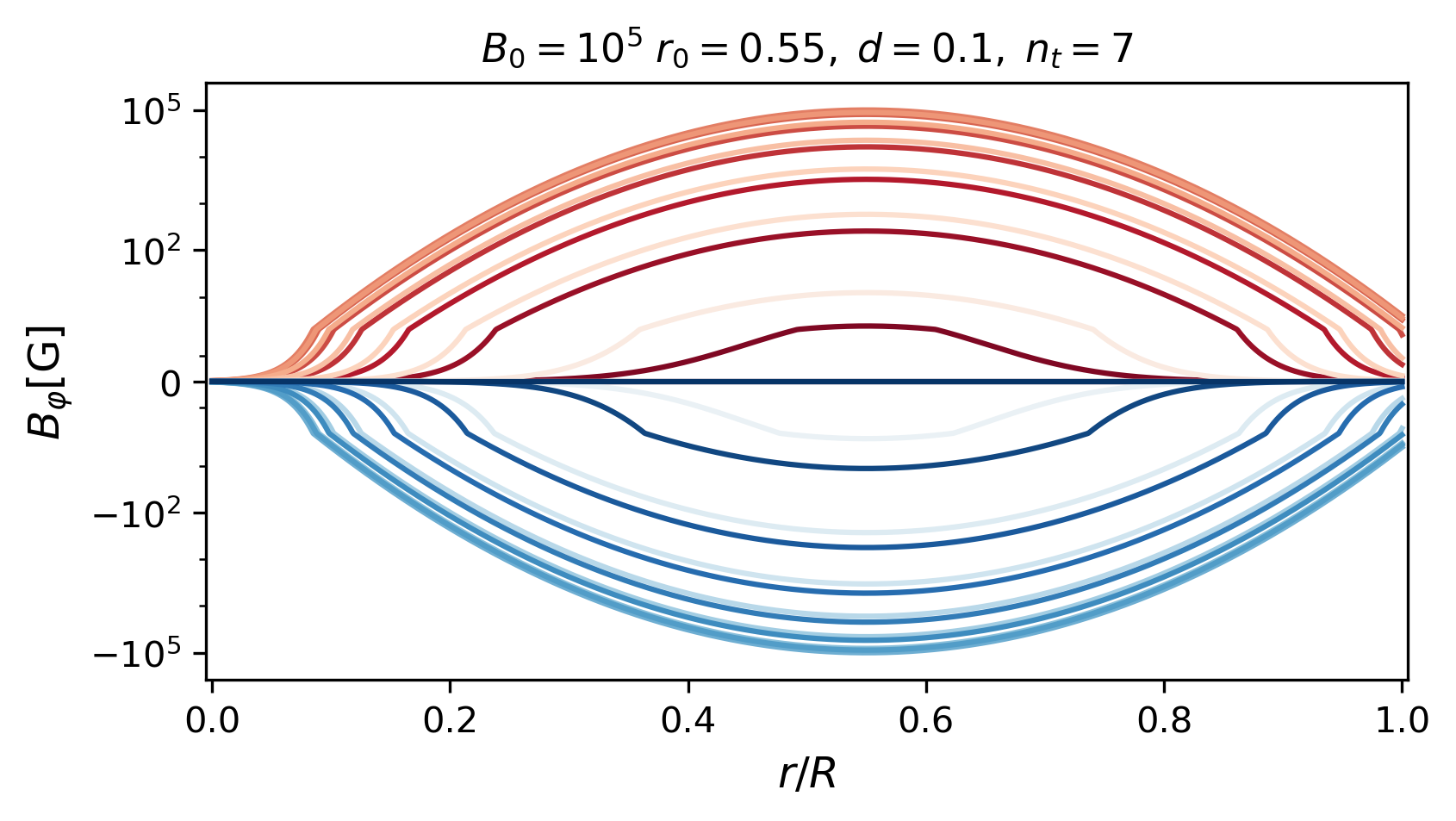}}
    \caption{Hemispheric toroidal magnetic field profile at $B_0=10^5\,\rm G$ with $r_0=0.55$, $d=0.1$, and $n_t=7$ as a function of the normalised radius at different colatitudes from the north pole (dark red) to the south pole (dark blue).}
    \label{fig:b_phi_hemispheric}
\end{figure}

\begin{figure*}
    \centering
     \resizebox{\hsize}{!}{\includegraphics{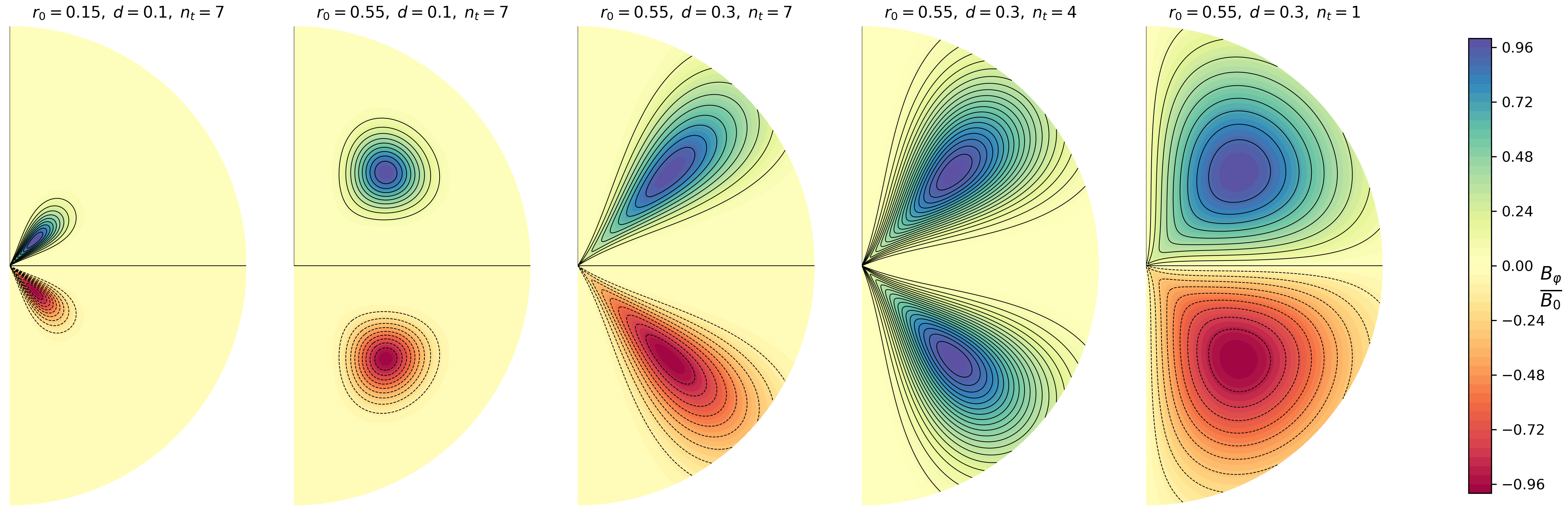}}
    \caption{Hemispheric toroidal  magnetic configuration (normalised by its amplitude) for different sets of parameters (the isolines are represented in black).}
    \label{fig:hemispheric_config}
\end{figure*}

Depending on the history of the differential rotation of the star, other magnetic topologies than the configuration adopted in the previous section
could emerge in the radiative zone. For instance, a radial differential rotation is likely to produce a strong toroidal field in the inner region of the radiative zone, whereas cylindrical differential rotation is expected to produce a strong toroidal field at the upper region of the stably stratified zone \citep{Jouveetal2020}. \cite{Zahnetal2007} found a toroidal field resulting from the interaction of the differential rotation of a solar-like tachocline and an initial dipolar poloidal fossil magnetic field. The resulting field will be composed of two magnetic distributions localised in the northern and southern hemispheres and vanishing at the equator.
Such magnetic configurations can be modelled by the following field distribution:
\begin{equation}
    B_\varphi (r, \theta) = B_0 \exp{\left[-\frac{\left(r-r_0\right)^2}{2d^2}\right]} \sin^{n_t}{\left(2\theta\right)},
\end{equation}
where $B_0$ is the magnetic field amplitude, $r_0$ is the radial position of the centre of distribution, $d$ is its radial extension, and $n_t$ parametrises its latitudinal extension. Figure\;\ref{fig:b_phi_hemispheric} shows the radial profile of such a field at different colatitudes for $B_0=10^5\,\rm G$, $r_0=0.55$, $d=0.1$, and $n_t=7$. To get a sense of each of these parameters, Fig.\;\ref{fig:hemispheric_config} represents the hemispheric field configurations for different sets of parameters to assess their influence. We can see that for odd values of $n_t$ we obtain an antisymmetric field with respect to the equator (positive in the northern hemisphere and negative in the southern one) whereas for even values of $n_t$ the field is symmetric (positive in the
two hemispheres). Increasing $n_t$ allows us to obtain two well-separated distributions in latitude. 
Using Eq.\;(\ref{eq:B_T}), we extract the Alfvén frequency, which depends on $r$ and $\theta$ in this case:
\begin{equation}
    \omega_{\rm A} (r,\theta) = \frac{1}{r \sin{\theta}}\frac{B_\varphi (r, \theta)}{\sqrt{\mu_0 \rho_0(r)}}.
\end{equation}
For antisymmetric fields, $\omega_{\rm A}$ is negative in the southern hemisphere, but this has no impact on the MLTO (Eq.\;\ref{eq:mlto_general}) since it is invariant with respect to the sign of the Alfvén frequency, as we discussed it in Sect.\; \ref{subsect:glte}.

To perform seismic diagnosis or any other applications such as AM transport or tidal dissipation by MGI waves in this more general case, we have to solve the general MLTE (Eq.\;\ref{eq:mlto_general2}). We use the representative stellar model of the typical $1.6{\rm M}_\odot$ $\gamma$\,Dor star rotating uniformly at the ZAMS as presented in Sect.\;\ref{sect:mesa_setup}.
To apply the magnetic TAR using this hemispheric configuration, we have to choose carefully the parameters that define the field and its amplitude because we have to verify 
the hierarchy of the frequencies imposed by the magnetic TAR for the entire domain of parameters, as well as avoid the magnetic critical layers (singularity when $\mathcal{A}(r,\theta)=0$). 
The magnetic field that we take should have an amplitude $B_0$ that does not exceed $ 10^5\;\rm G$ with a radial position $r_0$ smaller than 0.35, a maximal radial extension of 0.2, and a latitudinal extension sufficiently far from the equator ($n_t\ge 2$). 
\begin{figure}
    \centering
     \resizebox{\hsize}{!}{\includegraphics{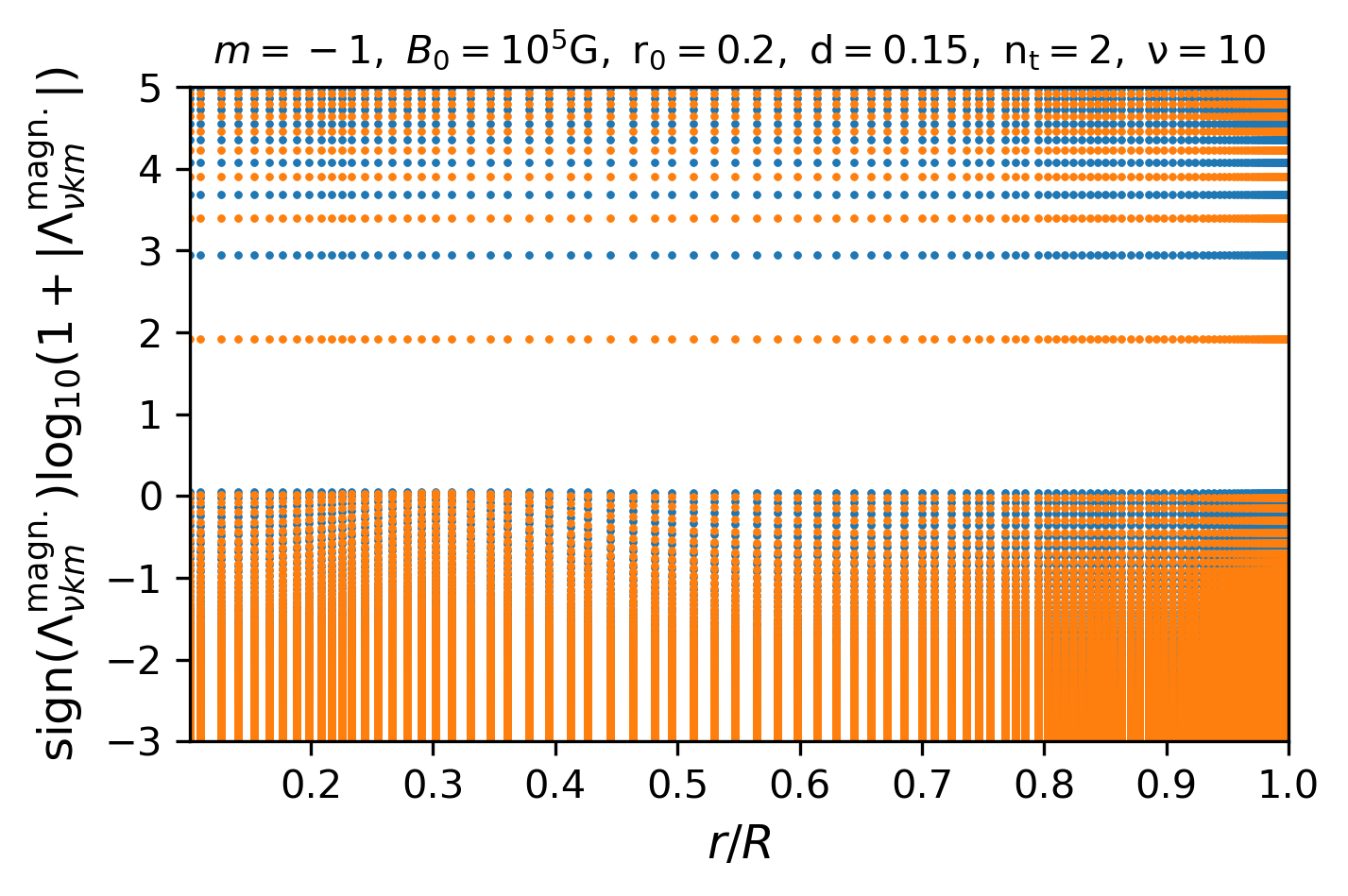}}
     \resizebox{\hsize}{!}{\includegraphics{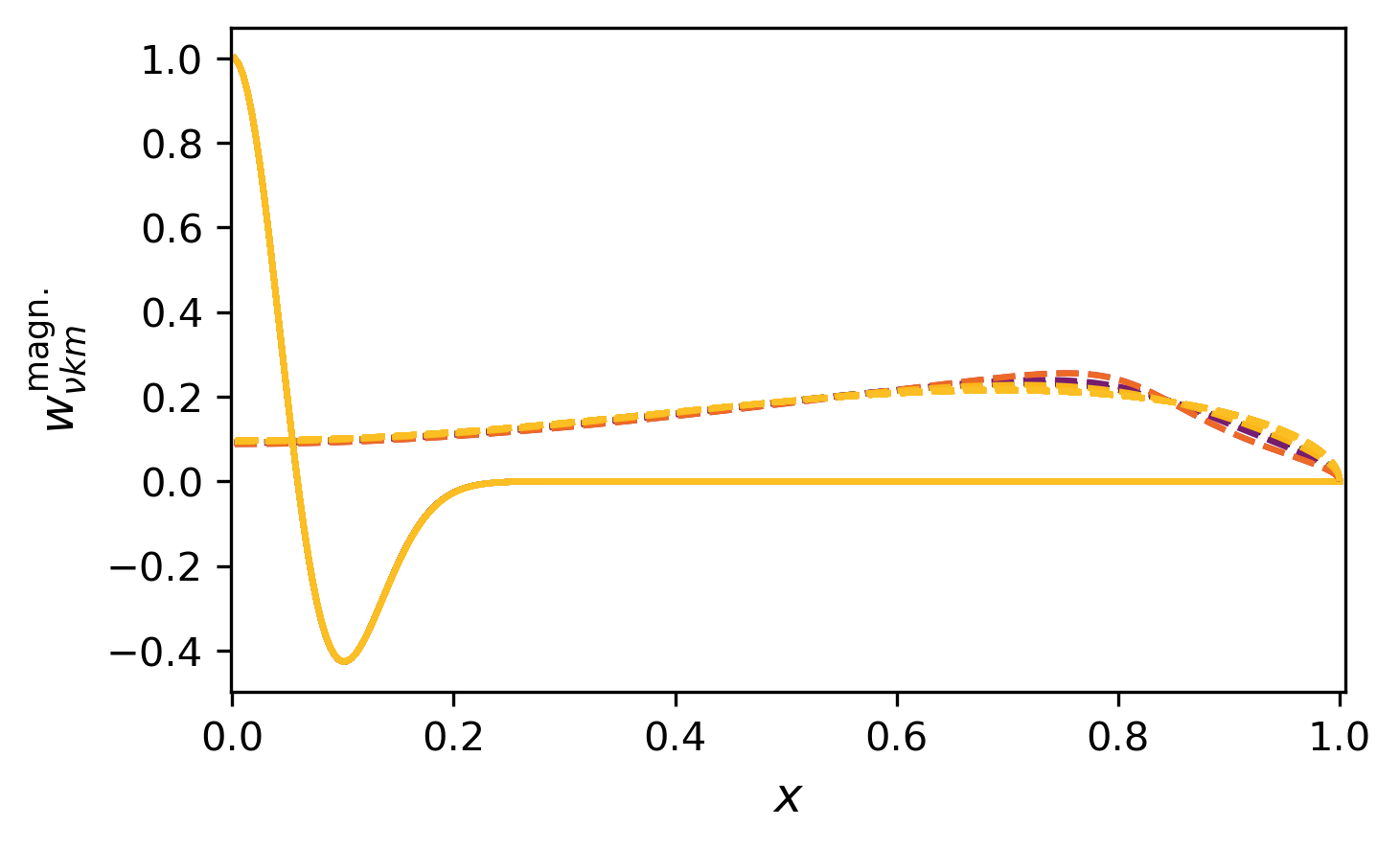}}
    \caption{Same as Fig.\;\ref{fig:eigenvalues} (top)    and Fig.\;\ref{fig:hough} (bottom) but for a magnetic hemispheric field with $B_0=10^5\;\rm G$, $r_0=0.2$, $d=0.15$, and $n_t=2$.}
    \label{fig:eigenvalues_hough_hemispheric}
\end{figure}

\begin{figure}
    \centering
     \resizebox{\hsize}{!}{\includegraphics{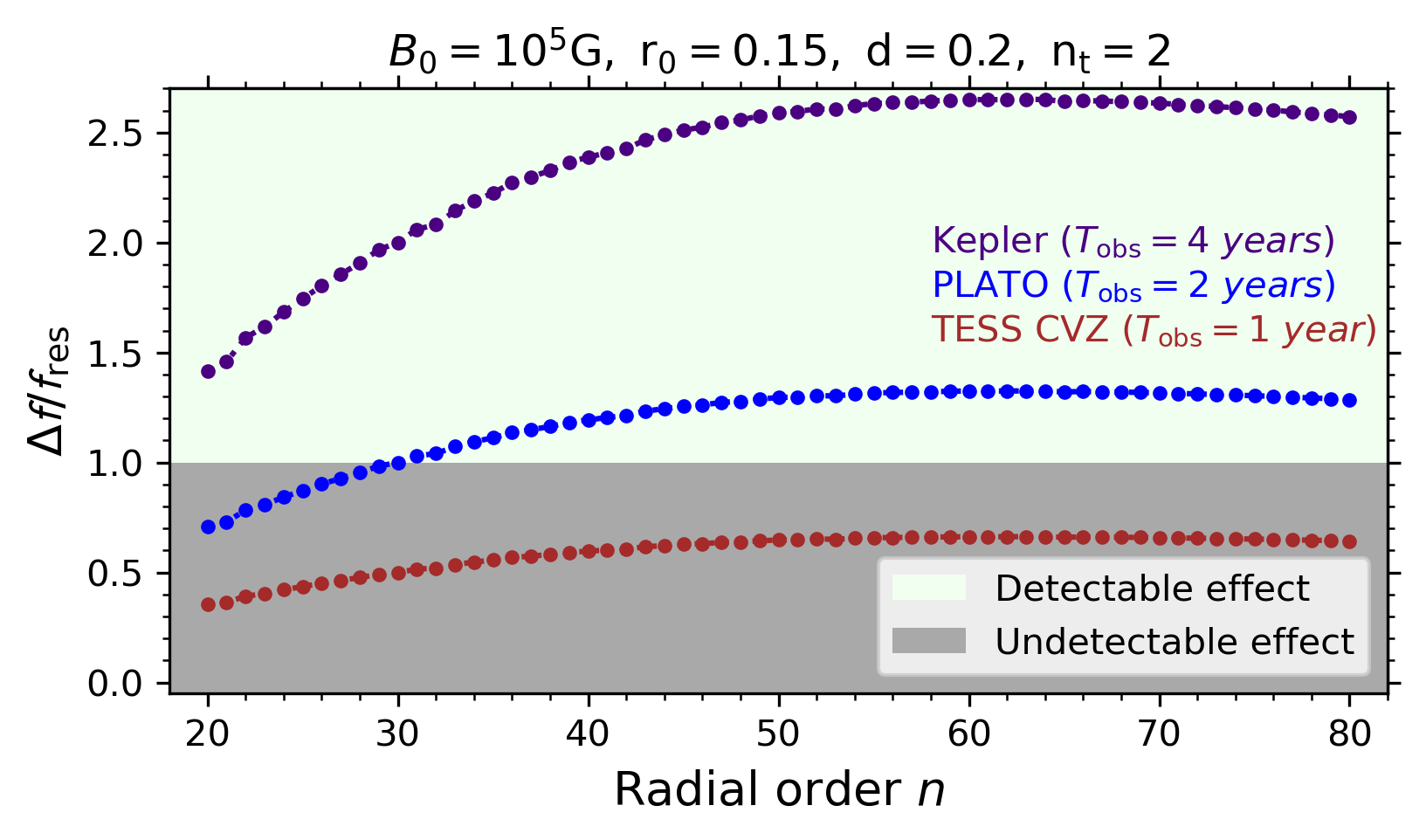}}
    \resizebox{\hsize}{!}{\includegraphics{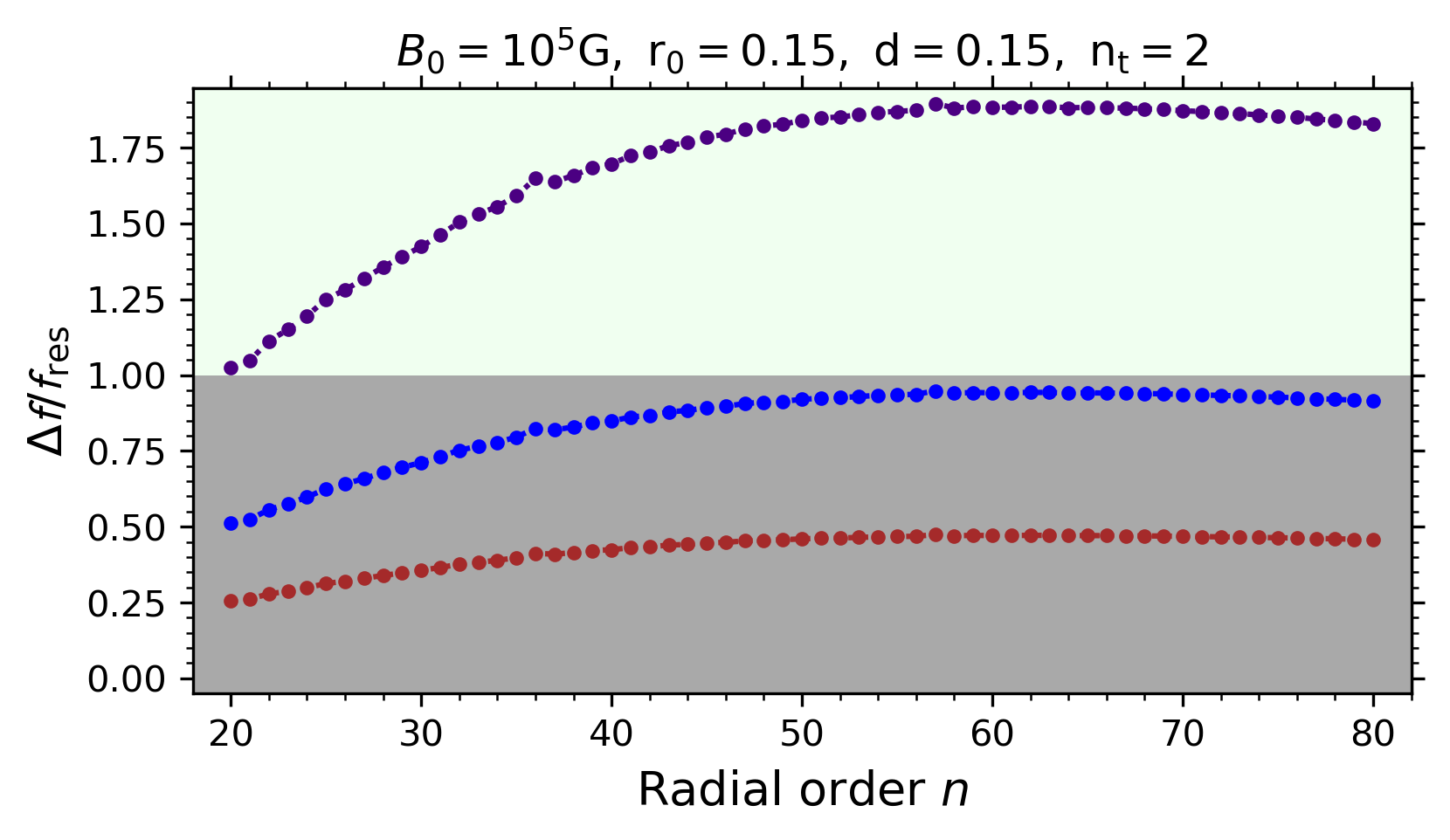}}
      \resizebox{\hsize}{!}{\includegraphics{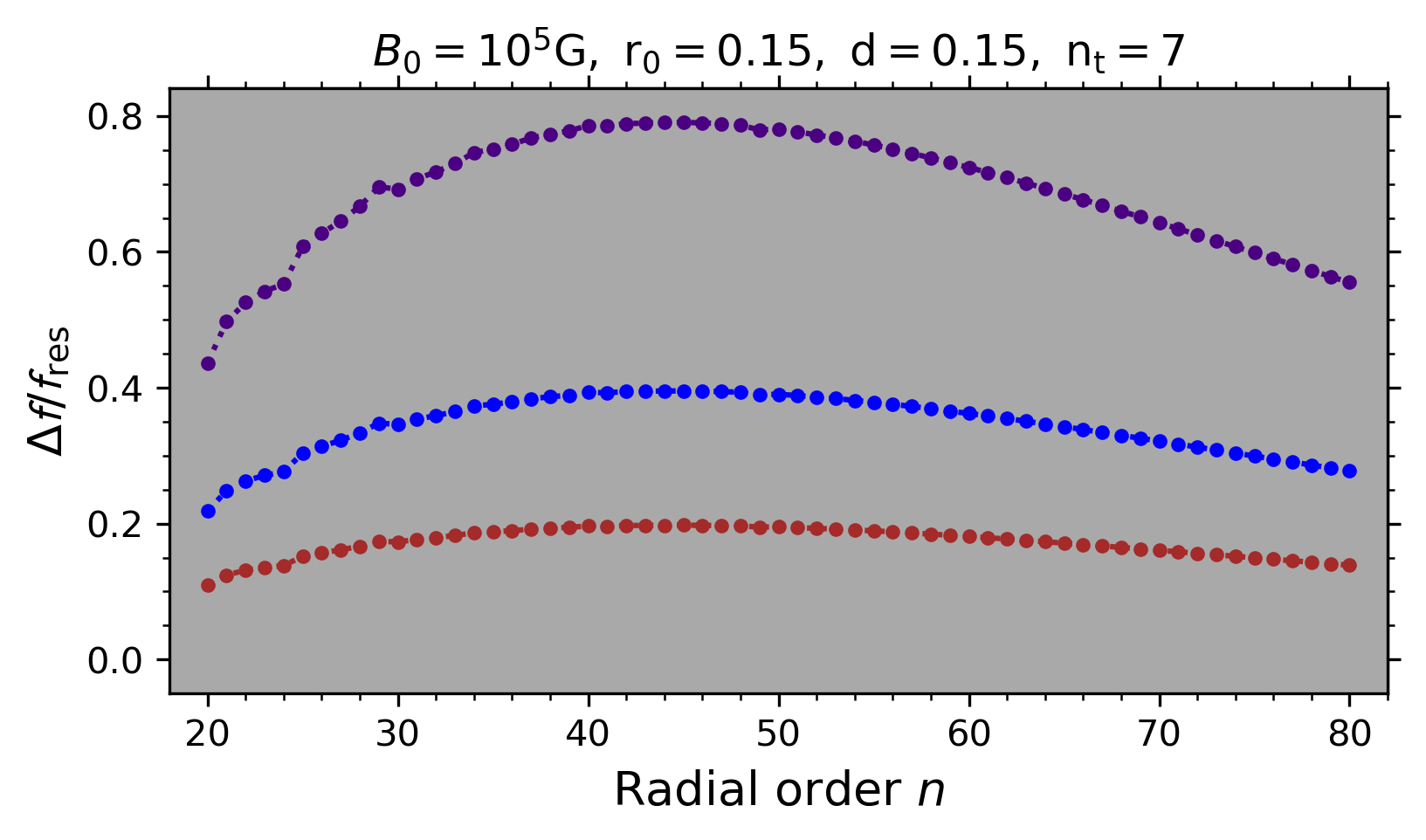}}
    \resizebox{\hsize}{!}{\includegraphics{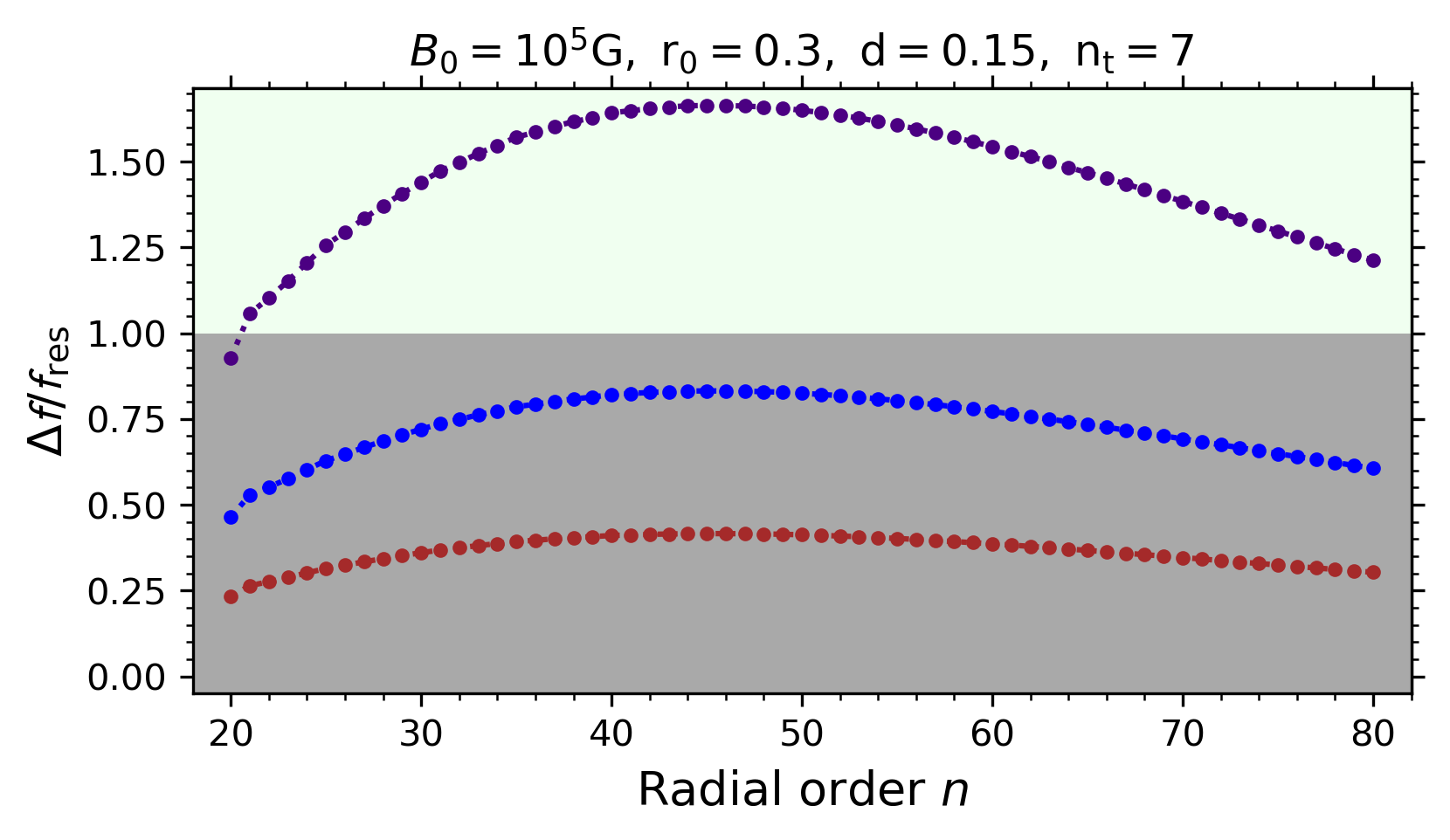}}
    \caption{Effect of the magnetic distribution on the detectability of the hemispheric magnetic field signature on the $\rm g$ modes $\{k=0,m=1\}$ as a function of the radial order  $n$ using the $1.6{\rm M}_\odot$ $\gamma\,$Dor model at ZAMS rotating at $ \Omega = 22.19\,\Omega_{\odot} $ and $B_0=10^5\,\rm G$.}
    \label{fig:detect_hemispheric}
\end{figure}

By solving the MLTE using the method described in Sect.\;\ref{sect:hough_eigen} within the validity of our formalism, we find that the eigenvalues and the Hough functions are hardly modified compared to the non-magnetic ones (Fig.\;\ref{fig:eigenvalues_hough_hemispheric}). As a consequence, the detectability of such a field is limited (within our validity domain). An important first conclusions is that the detectability is intrinsically linked  to the magnetic configuration. We show in Fig.\;\ref{fig:detect_hemispheric} the effect of the magnetic distribution on the detectability of the hemispheric magnetic field's signature on the $\rm g$\;modes $\{k=0,m=1\}$ for different sets of parameters. We find that the detectability increases with the radial position $r_0$ (third and fourth panels in Fig.\;\ref{fig:detect_hemispheric}) and extension $d$ (first and second panels in Fig.\;\ref{fig:detect_hemispheric}) of the distribution. This is caused by the fact that the Alfvén frequency reaches a higher maximum far from the centre because it is inversely proportional to the density which decreases rapidly towards the stellar surface. Moreover, increasing the parameter $n_t$ implies decreasing (second and third panels in Fig.\;\ref{fig:detect_hemispheric}) the latitudinal extension and reduces the detectability. Hence, the more the Alfvén frequency is localised, the less detectable it becomes.  
So even if hemispheric fields are present in intermediate-mass stars, the detectability of such complex magnetic fields is low, even in the most favourable cases. The magnetic topology considered in Sect.\;\ref{sec:field_topo} ($\tilde{b}_\varphi(r)\sin\theta$) is more favourable in terms of detectability because, in that case, the Alfvén frequency, $\omega_{\rm A}$, is distributed in all latitudes (independent of $\theta$), whereas here it is latitudinally localised. A simple physical interpretation follows: an equatorial field is always in the propagation region of MGI modes that can be equatorially trapped; this is not the case of hemispheric fields, which are potentially strongly localised at given latitudes. This conclusion is only valid within the scope of our validity domain. 

To examine more extended and stronger magnetic fields we should go beyond the magnetic TAR, in order not to limit to its restricted domain of validity. Other than the limitation of the TAR, another point to consider in future work is the role of the magnetic critical layers, resulting in dissipated waves. An adequate formalism and technique should be developed and adopted to tackle such a resonant phenomenon because it prevents the modes from occurring.

\section{Discussion and conclusions}\label{sect:conclusion}
In this work we investigate the effects of an axisymmetric, purely toroidal deep internal large-scale magnetic field of fossil origin or generated by a dynamo action on the oscillation frequencies of gravito-inertial modes.
We carry out a new generalisation of the standard TAR that abandons  the non-magnetic and the uniformly rotating assumptions. The magnetic TAR developed here takes the general differential rotation, $\Omega(r,\theta)$, and axisymmetric toroidal magnetic field, $B_\varphi(r,\theta)$, into account in a non-perturbative way.

As a first asteroseismic application of this general formalism, we apply the magnetic TAR to representative stellar models of typical $\gamma$\;Dor and SPB stars during 
their MS evolution. These gravity or gravito-inertial wave pulsators hold to potential to probe mixing processes \citep{Degroote2010, Pedersen2021}, rotation \citep{vanreeth2016,vanreeth2018, Ouazzani2017,Papics2017, Aerts2017, aerts2019,Ouazzani2020,Saio2021,Henneco2021, dhouib2021a, dhouib2021b}, and magnetic fields \citep{Prat2019,Prat2020, VanBeeck2020}.
We find that the toroidal magnetic field affects the theoretical description of gravito-inertial modes, forming MGI modes.
The Laplace tidal equation is altered and gets a radial dependence, propagating into solutions that are modified relative to those of the standard TAR. More precisely, the magnetic field implies a shift in the extrema of the Hough functions towards the equator of the star. 
For equatorial, initially fossil azimuthal magnetic fields, which can be regenerated through potential dynamo action, this effect becomes more apparent towards the inner radiative region since the magnetic field is the strongest at the convective--radiative interface and the weakest near the surface. The effect is much less discernible for the hemispheric fields since the Alfvén frequency is latitudinally localised. 
Indeed, since most $\rm g$ modes are trapped in the equatorial belt, the modes are only weakly impacted by a hemispheric field because it is localised far from the equator.

We find that a fossil magnetic field reduces the period spacing values for $\rm g$\;modes for all radial orders. For $\rm r$\;modes, the magnetic field increases the period spacings for low radial orders (long pulsation periods) and decreases for high radial orders (short pulsation periods). 
The mode frequency shifts are potentially detectable using high-precision space-based photometry for most of the modes when the magnetic field is sufficiently strong at the inner radiative region 
(of order $10^5$\,G).
The detectability of the toroidal magnetic effects using observations from modern space photometry decreases with increasing stellar age (decreasing $X_{\mathrm{c}}$) for $\gamma$\;Dor and SPB stars. 
Similar effects were found for a more general and complex magnetic configuration with an amplitude of the order of $10^5\,\rm G$. We find that the more the field is localised, especially at latitudes far from the equator, the smaller the signature of the magnetic field on the waves.
The amplitudes that we test here, within the validity domain of the magnetic TAR,  are in  agreement with the latest modelling and simulations of magnetic fields in the radiative zones of early-type stars 
\citep[e.g.][]{Fulleretal2019, Petitdemange2021}.

Even though a toroidal magnetic field is in principle detectable, it does not have a distinctive signature in the period spacing pattern. Rather, it introduces mode period shifts that occur simultaneously with those produced by other physical effects, such as  the centrifugal acceleration or differential rotation.
In addition, these effects are of the order of (or smaller than) the effects of other stellar physical processes in stellar models (e.g. rotational mixing, atomic diffusion, etc.). The degeneracy among these astrophysical phenomena can mask the magnetic effect in forward asteroseismic modelling analyses. In order to tackle this problem, work is being  \citep{vanreeth2018,dhouib2021a, dhouib2021b, Mombarg2022} and will be done to improve the modelling of these processes, with the aim to unravel and discern the different signatures. The inclusion of magnetic effects deduced from oscillation frequencies will provide important observational constraints on the theory of stellar interiors for early-type stars, which will subsequently lead to better magneto-asteroseismic modelling.

In addition to the asteroseismic application, 
our new magnetic TAR theory opens the way to various applications in stars and planets. One prominent example is the study of AM transport induced by low-frequency MGI waves in differentially rotating magnetic stars and planets. This allows us to generalise the prescription developed by \cite{mathis2009}, who studied the action of the differential rotation on gravito-inertial waves in non-magnetised radiative zones and its feedback on the AM transport. 
Furthermore, the magnetic TAR can be used to study the dissipation of stellar and planetary tides in rotating, magnetised, stably stratified regions. Stably stratified zones in giant planets can be magnetised \citep{Debras2019, Mankovich2021}, and as such the magnetic TAR now allows for the development of a new prescription of the dynamical tide in non-convective zones. In this way, we can go beyond the formalism developed by \cite{Ahuir2021}, who tackled this topic by assuming a non-rotating non-magnetised body. 
An additional possible next step of our work is to apply the magnetic TAR to simulated toroidal magnetic configurations \citep{Zahnetal2007, Jouveetal2020, Petitdemange2021} and compare its signature and detectability with respect to the distribution of the field.

Even though the toroidal configurations studied here are unstable, we chose not to consider a complex helical field geometry as a first step in constructing the magnetic TAR to unravel the signature of magnetic fields on MGI waves. This is the best road as we have shown that observational constraints on the strength of the toroidal field can be deduced from modern space data. Such an inference can then naturally guide theoretical follow-up work via the study of the feasibility of carrying out a new TAR generalisation with the inclusion of a poloidal magnetic field. This would lead to a general magnetic TAR that takes any mixed magnetic configurations into account  in a non-perturbative way while adhering to observed asteroseismic signals of stars.

Finally, another interesting perspective of this work is to go beyond the magnetic TAR as was done by \cite{Valade2018}, who used a Hamiltonian ray-tracing method, and \cite{Asai2015} \citep[see also][]{Asai2016,Lee2018}, who used general series expansion on spherical harmonics to explore the impact of different general magnetic field topologies on gravito-inertial waves. Their formalism can be adapted to perform asteroseismic modelling of observed stars.

\begin{acknowledgements}
    We thank the referee for her/his positive and constructive report, which has allowed us to improve the quality of our article. H.D. and S.M. acknowledge support from the CNES PLATO grant at CEA/DAp. TVR gratefully acknowledges support from the Research Foundation Flanders (FWO) under grant agreement No. 12ZB620N and V414021N. This research was supported in part by the National Science Foundation under Grant No. NSF PHY-1748958. CA is supported by the KU\,Leuven Research Council (grant C16/18/005: PARADISE) as well as from the BELgian federal Science Policy Office (BELSPO) through a PLATO PRODEX grant.
\end{acknowledgements}

\bibliographystyle{aa}
\bibliography{bibliography}

\begin{appendix}
\section{MESA controls inlist}
\label{sec:MESA_inlist}
In this appendix we report the control section of the MESA inlist used to compute the stellar evolution
models of the $1.6\mathrm{M}_\odot$ and $5\mathrm{M}_\odot$, $Z=0.02$ stars:

\begin{Verbatim}

&controls

  ! starting specifications
    initial_mass  = 1.6 (GammaDor) and 5 (SPB) ! in Msun units
    initial_z     = 0.02
    use_Type2_opacities = .true.
    Zbase        = 0.02

  ! stop criteria
    xa_central_lower_limit_species(1) = 'h1'
    xa_central_lower_limit(1) = 0.000001

  !—————————————————————————————————————— WIND
  cool_wind_RGB_scheme = 'Reimers'
  cool_wind_AGB_scheme = 'Blocker'
  RGB_to_AGB_wind_switch = 1d-4
  Reimers_scaling_factor = 0.2
  Blocker_scaling_factor = 0.5
  use_accreted_material_j = .true.
  accreted_material_j = 0

  !——————————————————————————————————————OVERSHOOTING
  overshoot_scheme(1) = 'exponential'
  overshoot_zone_type(1) = 'any'
  overshoot_zone_loc(1) = 'any'
  overshoot_bdy_loc(1) = 'any'
  overshoot_f(1) = 0.015
  overshoot_f0(1) = 0.004

  !—————————————————————————————————————— MESH
  mesh_delta_coeff = 0.7
  varcontrol_target = 0.7d-3
  predictive_mix(1) = .true.
  predictive_superad_thresh(1) = 0.005
  predictive_avoid_reversal(1) = 'he4'
  predictive_zone_type(1) = 'any'
  predictive_zone_loc(1) = 'core'
  predictive_bdy_loc(1) = 'top'
  dX_div_X_limit_min_X = 1d-4
  dX_div_X_limit = 5d-1
  dX_nuc_drop_min_X_limit = 1d-4
  dX_nuc_drop_limit = 1d-2

/ ! end of controls namelist


\end{Verbatim}

\section{Additional tables} \label{sec:tables}
\begin{landscape}
\begin{table}
\caption{Detectability of the signature of toroidal magnetic fields in the modelled $1.6{\rm M}_\odot$ $\gamma$\,Dor star model during MS evolution for three modes at different rotation rates, $\Omega$, and magnetic field amplitude, $B_0$. We also list the ranges of radial orders, $n$, and spin parameters, $\nu$.}
\label{table:detect_gamma}
\begin{tabular}{c c | c | c c | c c c c c c c c c c  c c}
\hline\hline
\multicolumn{2}{c|}{mode} & Evolution phase & \multicolumn{2}{c|}{$\Omega$} & $B_0$ & \multicolumn{2}{c}{$n$} & \multicolumn{2}{c}{$\nu$} & $\Delta f_{\rm max}$ &$n_{\mathrm{detect},min}$ & $\Delta f_{\rm max} /f_{\rm res} $& $n_{\mathrm{detect},min}$ & $\Delta f_{\rm max} /f_{\rm res} $& $n_{\mathrm{detect},min}$ & $\Delta f_{\rm max} /f_{\rm res} $\\ $k$ & $m$ & & [$\Omega_{\odot}$] & [c/d] & $[\rm G]$ & (min) & (max) & (min) & (max)& [c/d] &\multicolumn{2}{c}{(TESS)}& \multicolumn{2}{c}{(PLATO)}& \multicolumn{2}{c}{(\textit{Kepler})}\\
\hline
0&1     & near   & 22.19 & 0.79 & $6\times10^{4}$ & 20 & 80  & 1.54 & 6.7 & 0.0015 & \ldots & 0.57 & 73 & 1.11 & 37 & 2.23\\
   &   &    ZAMS   &   &  & $2\times10^{5}$          & 20 & 80    &  1.54 & 6.2 & 0.02 & 20  & 7.54   & 20    & 15.1 & 20 & 30.19 \\ 
             &   &      &  &  & $6\times10^{5}$   & 20 & 37    & 1.48 & 2.49 & 0.11 &  20  & 39.2  & 20    & 78.43 & 20 & 156.87 \\ 
                \cline{4-17} 
                &  &  &  15.56 & 0.56 & $6\times10^{4}$ &  20 & 80 & 1.04 & 4.62 & 0.0015 &  \ldots & 0.55 & 73 & 1.1 & 35 & 2.21\\
              &  &  &   & & $2\times10^{5}$ & 20 & 80 & 1.04 & 4.29 & 0.02 & 20 & 7.39 & 20& 14.78 & 20 & 29.57\\
              &  &  &       & &  $6\times10^{5}$ & 20 & 37 & 0.99 & 1.69 & 0.11 & 20 & 41.35 & 20 & 82.72 & 20 & 165.44\\
                \cline{3-17} 
                &  & mid-MS & 15.56 &  0.56 &  $6\times10^{4}$          & 20 & 80    & 0.97 & 4.32 & 0.0009 & \ldots  & 0.34  & \ldots    & 0.67 & 60 & 1.34 \\
              &  &  &  &  & $2\times10^{5}$          & 20 & 80    & 0.96 & 4.16 & 0.01 & 20  & 3.87  & 20    & 7.74 & 20 & 15.48 \\
              &  &       &  &  & $6\times10^{5}$   & 20 & 76    & 0.94 & 2.89 & 0.11 & 20  & 41.72     & 20    & 83.44  & 20 & 166.88   \\ 
                \hline
                0&2     & near   &  22.19   & 0.79 & $6\times10^{4}$ & 20 & 80 & 0.8 & 3.35 & 0.003 &  73 & 1.12 & 37 & 2.23 & 20 & 4.46      \\ 
&    & ZAMS  &    & & $2\times10^{5}$          & 20 & 80    & 0.79 & 3.1 & 0.04 & 20  & 15.12   & 20    & 30.23 & 20 & 60.47\\ 
               & &       &  &  & $6\times10^{5}$   & 20 & 37    & 0.76 & 1.25 & 0.21 & 20  & 79.6  & 20    & 159.2 & 20 & 318.41 \\ 
                \cline{4-17} 
               &  &  &  15.56 & 0.56 & $6\times10^{4}$  & 20 & 80 & 0.55 & 2.32 & 0.003 &  73 & 1.1 & 36 & 2.21 & 20 & 4.43\\
              &  &  &   &  & $2\times10^{5}$  & 20 & 80 & 0.54 & 2.15 & 0.04 & 20 & 14.84 & 20 & 29.67 & 20 & 59.35 \\
              &  &  &       &  & $6\times10^{5}$ & 20 & 36 & 0.52 & 0.85 & 0.21 & 20 & 76.36 & 20 & 152.73 & 20 & 305.45
               \\ \cline{3-17} 
               & & mid-MS &  15.56 & 0.56 & $6\times10^{4}$  & 20 & 80    & 0.51   & 2.17 & 0.0018 &\ldots & 0.67   & 60    & 1.35   & 29 &  2.69   \\
               & &  &   &  & $2\times10^{5}$          & 20 & 80    & 0.51   & 2.09 & 0.02 & 20 & 7.77   & 20    & 15.54   & 20 & 31.08     \\
               & &     &   &  & $6\times10^{5}$   & 20 & 78    & 0.5   & 1.46 & 0.24 & 20 & 86.81 & 20    & 173.62 & 20 & 347.24 \\ 
                \hline
-2&-1  & near   &  22.19 & 0.79  & $6\times10^{4}$   & 20 & 80    & 7.51  & 19.75 & 0.006 & 20    & 2.15   & 20    & 4.29 & 20 & 8.58\\ 
                \cline{4-17} 
              &  & ZAMS &  15.56 & 0.56 & $6\times10^{4}$ & 20 & 80 & 6.66 & 14.03 & 0.006 & 20 & 2.23 & 20 & 4.47 & 20 & 8.93 \\
                \cline{3-17} 
              &  & mid-MS &  15.56 & 0.56 & $6\times10^{4}$  & 20 & 80 & 6.61 & 13.71 & 0.003 & 20 & 1.25 & 20 & 2.51 & 20 & 5.02 \\
              &  &       &  &  & $2\times10^{5}$ & 22 & 45 & 6.06 & 7.51 & 0.027 & 22 & 9.99 & 22 & 19.98 & 20 & 39.96\\ 
                \hline
\end{tabular}
\end{table}
\end{landscape}

\begin{landscape}
\begin{table}
\caption{Same as Table\;\ref{table:detect_gamma}, but for the modelled $5{\rm M}_\odot$ SPB star.}
\label{table:detect_spb}
\centering
\begin{tabular}{c c | c | c c | c c c c c c c c c c  c c}
\hline\hline
\multicolumn{2}{c|}{mode} & Evolution phase & \multicolumn{2}{c|}{$\Omega$} & $B_0$ & \multicolumn{2}{c}{$n$} & \multicolumn{2}{c}{$\nu$} & $\Delta f_{\rm max}$ &$n_{\mathrm{detect},min}$ & $\Delta f_{\rm max} /f_{\rm res} $& $n_{\mathrm{detect},min}$ & $\Delta f_{\rm max} /f_{\rm res} $& $n_{\mathrm{detect},min}$ & $\Delta f_{\rm max} /f_{\rm res} $\\ $k$ & $m$ & & [$\Omega_{\odot}$] & [c/d] & $[\rm G]$ & (min) & (max) & (min) & (max)& [c/d] &\multicolumn{2}{c}{(TESS)}& \multicolumn{2}{c}{(PLATO)}& \multicolumn{2}{c}{(\textit{Kepler})}\\
\hline
0 & 1     & near   & 11.38 & 0.41 & $4\times10^{4}$ & 20 & 80  & 1.91 & 8.18 & 0.0005 & \ldots & 0.18 & \ldots & 0.35 & \ldots & 0.71\\
  &  & ZAMS   &  &  & $10^{5}$ & 20 & 80  & 1.91 & 7.95 & 0.003 & 69 & 1.22 & 37 & 2.43 & 20 & 4.86\\
 &      &    &  &  & $5\times10^{5}$ & 20 & 27  & 1.81 & 2.36 & 0.036 & 20 & 13.05 & 20 & 26.1 & 20 & 52.2   \\
  \cline{4-17}
  &     &    & 3.94 & 0.14 & $10^{5}$ & 20 & 80  & 0.59 & 2.6 & 0.003 & 68 & 1.19 & 31 & 2.39 & 20 & 4.77  \\
    &     &    &  &  & $5\times10^{5}$ & 20 & 24 & 0.56 & 0.65 & 0.042 & 20 & 15.29 & 20 & 30.59 & 20 & 61.17 \\
    \cline{4-17}
   &     &    & 1.18 & 0.04 & $10^{5}$ & 20 & 80  & 0.16 & 0.68 & 0.004 & 54 & 1.51 &  26 & 3.02 & 20 & 6.05 \\
      &     &    &  &  & $5\times10^{5}$ & 20 & 23  & 0.15 & 0.17 & 0.047 & 2à & 17.17 &  20 & 34.35 & 20 & 68.69 \\
 \cline{3-17} 
 &      & mid-MS  & 3.94 & 0.14 & $10^{5}$ & 20 & 80  & 0.54 & 2.41 & 0.0015 & \ldots & 0.55 & 73 & 1.09 & 31 & 2.18 \\
  &      &  &  &  & $5\times10^{5}$ & 20 & 59 & 0.53 & 1.45 & 0.033 & 20 & 11.95 & 20 & 23.89 & 20 & 47.78   \\
  \cline{4-17}
  &      &   & 1.18 & 0.04 & $10^{5}$ & 20 & 80  & 0.15 & 0.63 & 0.0019 & \ldots & 0.68 & 57 & 1.36 & 27 & 2.71\\
    &      &   &  &  & $5\times10^{5}$ & 20 & 50  & 0.15 & 0.33 & 0.036 & 20 & 13.33 & 20 & 26.67 & 20 & 53.34\\
   \cline{3-17}
  &      & near   & 1.18 & 0.04 & $10^{5}$ & 20 & 80 &  0.04 & 0.17 & 0.008& 27 & 3.07 & 20 & 6.14 & 20 & 12.3 \\
    &      & TAMS  &  & & $5\times10^{5}$ & 20 & 44 &  0.04 & 0.07 & 0.23 & 20 & 85.96 & 20 & 171.91 & 20 & 343.82 \\
  
\hline
0 & 2     & near   & 11.38 & 0.41 & $10^{5}$ & 20 & 80  & 0.97 & 3.98&  0.007 & 37 & 2.44 & 20 & 4.88 & 20 & 9.75\\
 &     & ZAMS   &  &  & $5\times10^{5}$ & 20 & 28 & 0.93 & 1.22 & 0.079 & 20 & 28.75 & 20 & 57.51 & 20 & 115.01 \\
  \cline{4-17}
 &     &    & 3.94 & 0.14 & $10^{5}$ & 20 & 80  & 0.32 & 1.31 & 0.0065 & 35 & 2.39 & 20 & 4.77 & 20 &9.54 \\
  &     &    &  &  & $5\times10^{5}$ & 20 & 24  & 0.3 & 0.35 & 0.066 & 20 & 24.19 & 20 & 48.38 & 20 & 96.75 \\
    \cline{4-17}
&      &   & 1.18 & 0.04 & $10^{5}$ & 20 & 80  & 0.09 & 0.37 & 0.007 & 33 & 2.6 & 20 & 5.19 & 20 & 10.38 \\
&      &   &  &  & $5\times10^{5}$ & 20 &  25 & 0.09 & 0.1 & 0.078 & 20 & 28.6 & 20 & 57.2 & 20 & 114.42\\
 \cline{3-17} 
 &      & mid-MS  & 3.94 & 0.14 & $10^{5}$ & 20 & 80  & 0.29 & 1.22 & 0.0031 & 74 & 1.08 & 36 & 2.17 & 20 & 4.33\\
  &      &  &  &  & $5\times10^{5}$ & 20 & 62 & 0.29 & 0.77 & 0.066 & 20 & 24.13 & 20 & 48.27 & 20 & 96.55 \\
  \cline{4-17}
   &      &   & 1.18 & 0.04 & $10^{5}$ & 20 & 80  & 0.09 & 0.34 & 0.0032 & 69 & 1.17 & 34 & 2.34 & 20 & 4.68 \\
      &      &   &  &  & $5\times10^{5}$ & 20 & 53  & 0.08 & 0.19 & 0.061 & 20 & 22.4 & 20 & 44.79 & 20 & 89.58 \\
   \cline{3-17}
  &      & near   & 1.18 & 0.04 & $10^{5}$ & 20 & 80 & 0.02 & 0.09 & 0.013 & 20 & 4.82 & 20 & 9.64 & 20 & 19.27\\  
    &      &  TAMS  &  &  & $5\times10^{5}$ & 20 &52 & 0.02 & 0.05 & 0.459 & 20 & 167.5 & 20& 335 &20 & 670.02 \\  
\hline
-2 & -1     & near   & 11.38 & 0.41 & $4\times10^{4}$ & 20 & 80 & 8.24 & 24.28 & 0.0018 & \ldots & 0.67 & 64 & 1.34 & 20 & 2.68 \\
 &      & ZAMS   &  & & $10^{5}$ & 20 & 31  & 7.93 & 10.18 & 0.005 & 20 & 1.97 & 20 &3.94 & 20 & 7.88\\
  \cline{4-17}
 &      &    & 3.94 & 0.14 & $4\times10^{4}$ & 20 & 79 & 6.08 & 9.15 & 0.0023 & \ldots & 0.86 & 20 & 1.71 & 20 & 3.43 \\
  &      &    &  &  & $10^{5}$ & \ldots & \ldots & \ldots & \ldots & \ldots & \ldots & \ldots & \ldots & \ldots & \ldots & \ldots \\
   \cline{4-17}
 &      &    & 1.18 & 0.04 & $4\times10^{4}$ & \ldots & \ldots & \ldots & \ldots & \ldots & \ldots & \ldots & \ldots & \ldots & \ldots & \ldots\\
  \cline{3-17}
   &      & mid-MS  & 3.94 & 0.14 & $4\times10^{4}$ & 20 & 80  & 6.1 & 9 & 0.0012 & \ldots & 0.45 & \ldots & 0.9 & \ldots & 1.8\\
 &      &   &  &  & $10^{5}$ & 40 & 63  & 6.1 & 6.98 & 0.006 & 20 & 2.28 & 20 & 4.55 & 20 & 9.1\\
 \cline{4-17}
    &      &   & 1.18 & 0.04 & $4\times10^{4}$ & \ldots & \ldots & \ldots & \ldots & \ldots & \ldots & \ldots & \ldots & \ldots & \ldots & \ldots \\
 \cline{3-17}
  &      & near TAMS  & 1.18 & 0.04 & $4\times10^{4}$ & \ldots & \ldots & \ldots & \ldots & \ldots & \ldots & \ldots & \ldots& \ldots & \ldots & \ldots  \\    
  
\end{tabular}
\end{table}
\end{landscape}

\end{appendix}
\end{document}